\documentclass[a4paper,fleqn,usenatbib]{mnras}
\usepackage[T1]{fontenc}
\usepackage{ae,aecompl}
\usepackage{graphicx}
\usepackage{amsmath}    
\usepackage{amssymb}    
\usepackage{mathptmx}
\usepackage{txfonts}

\newcommand{\etal}{{\it et al.}}

\title[Prediction for the second maximum of Solar Cycle 25] 
{Prediction  for the  amplitude  and second maximum of Solar Cycle 25
  and a comparison of the predictions based on
 strength of polar magnetic field and low latitude sunspot area} 

\author[J. Javaraiah]{J. Javaraiah\thanks{E-mail: jajj55@yahoo.co.in; jdotjavaraiah@gmail.com; jj@iiap.res.in}
\thanks{Formerly worked in Indian Institute of Astrophysics, Bengaluru-560 034,
India}\\
 Bikasipura, BSK 5th Stage, Bengaluru-560 111, India}

\date{Accepted XXX. Received YYY; in original form ZZZ}

\pubyear{2022}

\begin{document}
\label{fristpage}
\pagerange{\pageref{firstpage}--\pageref{lastpage}}
\maketitle

\begin{abstract}
The maximum  of  a  solar cycle contain two or more peaks, known as Gnevyshev
 peaks. Studies of this property of solar cycles may help for better
understanding the solar dynamo mechanism. We analysed the 13-month smoothed
 monthly mean Version-2 international sunspot number (SN) during the
 period 1874\,--\,2017
and found that there exists a good correlation between the amplitude
 (value of the main and highest peak) and the value of the second maximum
(value of the second highest peak) during the maximum of a solar cycle.
Using  this relationship and the earlier predicted value $86\pm18$ ($92\pm11$)
 of the amplitude  of Solar Cycle~25, here we predict a  value 
$73\pm15$ ($79\pm15$) for the second maximum of Solar Cycle~25. The 
ratio of the predicted second maximum to the amplitude is found to be 
 0.85, almost the same as that of Solar Cycle~24. The least-square cosine fits
 to the values of the peaks that occurred first and  second
during the maxima  of  Solar Cycles 12\,--\,24 suggest that in Solar Cycle~25
the second maximum would occur before the main maximum, the same as in
 Solar Cycle~24. However, these fits suggest $\approx$106 and  $\approx$119
 for the second maximum and the amplitude of Solar Cycle~25, respectively. 
Earlier, we analysed the combined  Greenwich and Debrecen  sunspot-group data
during 1874\,--\,2017 and predicted the amplitude of Solar Cycle~25 from  
the activity just after the maximum of Solar Cycle~24 in the  equatorial 
latitudes of the Sun's southern hemisphere.
Here from  the  hindsight of the results  we  found 
the earlier prediction is reasonably reliable.  We analysed the  polar-fields 
 data measured  in Wilcox Observatory  during Solar Cycles 20\,--\,24 and 
obtained a value $125\pm7$ for the amplitude of Solar Cycle 25. This is
 slightly larger--whereas the value $\approx$86 ($\approx$92) predicted from 
the activity in the  equatorial latitudes is slightly smaller--than the observed
 amplitude of Solar Cycle~24. This difference is discussed briefly.
 \end{abstract}

\begin{table*}
\centering
\caption{$R_{\rm M}$ represents the maximum (the largest 13-month smoothed
monthly mean SN) and $T_{\rm M}$  is the
corresponding  epoch (year) of a Solar Cycle $n$.
  $A_{\rm W}$  represents the
  13-month smoothed monthly mean areas
(msh) of the sunspot groups in the Sun's  whole-sphere
 at  $T_{\rm M}$  of a solar cycle.
 $\sigma_{\rm R}$  and $\sigma_{\rm W}$ 
represent  the errors in $R_{\rm M}$ and $A_{\rm W}$, respectively.
 $A^*_{\rm R}$ and    $A^*_{\rm W}$
represent  the sums of the areas (msh) of the sunspot groups
(normalized by 1000) in  $0^\circ-10^\circ$ latitude intervals of the
southern hemisphere  during  the time intervals $T^*_{\rm M}$ and 
 $T^*_{\rm W}$,   respectively,
just after $T_{\rm M}$ of a solar cycle.}
\label{table1}
\begin{tabular}{lccccccccc}
\hline
  \noalign{\smallskip}
$n$&$T_{\rm M}$&$R_{\rm M}$& $\sigma_{\rm R}$&$T^*_{\rm M}$&$A^*_{\rm R}$&
$A_{\rm W}$&$\sigma_{\rm W}$&$T^*_{\rm W}$&$A^*_{\rm W}$\\
  \noalign{\smallskip}
\hline
  \noalign{\smallskip}
12&1883.958&  124.4&12.5&1885.11-1885.71&  30.91& 1371&122&1884.86-1885.76&  50.94\\
13&1894.042&  146.5&10.8&1895.19-1895.70&  27.02& 1616&110&1894.94-1895.84&  35.04\\
14&1906.123&  107.1& 9.2&1907.27-1907.87&  32.34& 1043&139&1907.02-1907.92&  39.17\\
15&1917.623&  175.7&11.8&1918.77-1919.37&  32.48& 1535&170&1918.52-1919.42&  46.63\\
16&1928.290&  130.2&10.2&1929.44-1930.04&  70.20& 1324&123&1929.19-1930.09&  97.75\\
17&1937.288&  198.6&12.6&1938.44-1939.04&  71.62& 2119&176&1938.19-1939.09& 104.53\\
18&1947.371&  218.7&10.3&1948.52-1949.12& 103.85& 2641&210&1948.27-1949.17& 144.29\\
19&1958.204&  285.0&11.3&1959.35-1959.95&  31.67& 3441&208&1959.10-1960.00&  47.92\\
20&1968.874&  156.6& 8.4&1970.02-1970.62&  72.58& 1556&82&1969.77-1970.67&  80.58\\
21&1979.958&  232.9&10.2&1981.11-1981.71&  81.31& 2121&162&1980.86-1981.76& 104.26\\
22&1989.874&  212.5&12.7&1991.02-1991.62&  55.36& 2298&193&1990.77-1991.67&  86.67\\
23&2001.874&  180.3&10.8&2003.02-2003.62&  30.50& 2157&206&2002.77-2003.67&  47.62\\
24&2014.288&  116.4& 8.2&2015.44-2016.04&   6.20& 1560&116&2015.19-2016.09&  15.85\\
\hline
  \noalign{\smallskip}

\end{tabular}
\end{table*}

\begin{table*}
\caption[]{Hindsight: The values of intercept ($C$)  
and slope ($D$) of the linear
relationship between $A^*_{\rm R}$ of Solar Cycle $n$  and
 $R_{\rm M}$ of Solar Cycle $n+1$, and between
 $A^*_{\rm W}$ of Solar Cycle $n$ and $A_{\rm W}$ of 
Solar Cycle $n+1$,  that yielded the predictions  
for $R_{\rm M}$ and $A_{\rm W}$ of Solar Cycle $n+1$.
The corresponding values of the correlation coefficient ($r$),
Student's t ($\tau$), probability ($P$), number of data points ($N$),
and predicted value  are also given.}
\label{table2}
\begin{tabular}{lccccccccc}
\hline
  \noalign{\smallskip}
&&\multicolumn{4}{c}{$A^*_{\rm R} (n)$\,--\,$R_{\rm M} (n+1)$ relationship}\\ 
$n+1$ & $C$ & $D$& $r$ & $\tau$& $P$&  $N$ & Pred. value\\
18&$ 81.86\pm13.11$&$1.73\pm 0.34$&0.78& 2.18&$5.9\times10^{-2}$& 5 &$206.2\pm 20.5$\\
19&$ 76.19\pm10.78$&$1.92\pm 0.23$&0.87& 3.58&$1.1\times10^{-2}$& 6 &$275.7\pm 19.2$\\
20&$72.94 \pm8.59 $&$2.01\pm0.15 $&0.95&6.62 &$5.9\times10^{-4}$& 7 &$136.6 \pm18.0$\\
21&$80.90 \pm7.67$&$1.92\pm0.14$&0.94&6.95 &$2.2\times10^{-4}$& 8 &$220.3 \pm17.6 $\\
22&$79.78 \pm7.60$&$1.97\pm0.14$&0.95&7.75 &$5.5\times10^{-5}$& 9 &$240.1 \pm17.1 $\\
23&$82.28 \pm7.45$&$1.89\pm0.13$&0.94&7.56 &$3.2\times10^{-5}$& 10 &$186.8 \pm17.9 $\\
24&$81.87 \pm7.46$&$1.88\pm0.13$&0.94&7.98 &$1.1\times10^{-5}$& 11 &$139.4 \pm17.1 $\\
25&$74.04 \pm6.77$&$1.98\pm0.12$&0.94&8.45 &$3.5\times10^{-6}$& 12 &$86.3 \pm17.7$\\
\\
&&\multicolumn{4}{c}{$A^*_{\rm W} (n)$\,--\,$A_{\rm W} (n+1)$ relationship}\\ 
18&$720.32\pm181.02$&$14.99\pm3.34$&0.90&3.56&$1.9\times10^{-2}$& 5 &$2287\pm157$\\
19&$601.97\pm154,38$&$17.64\pm2.59$&0.94&5.53&$2.6\times10^{-3}$& 6 &$3147\pm180$\\
20&$515.25\pm123.12$&$19.39\pm1.77$&0.97&9.49&$1.1\times10^{-4}$& 7 &$1445\pm178$\\
21&$577.54\pm108.93$&$18.93\pm1.72$&0.97&10.35&$2.4\times10^{-5}$&8 &$2103\pm171$\\
22&$576.50\pm108.45$&$18.97\pm1.67$&0.97 &11.22&$5.0\times10^{-6}$ & 9 &$2555\pm161 $\\
23&$611.17\pm105.36$&$18.17\pm1.56$&0.97&10.55&$2.7\times10^{-6}$&10&$2186\pm175$\\
24&$612.37\pm104.99$&$18.13\pm1.53$&0.97&11.22&$7.1\times10^{-7}$&11&$1476\pm167$\\
25&$643.88\pm100.03$&$17.86\pm1.51$&0.96&11.63&$1.8\times10^{-7}$&12&$927\pm165$\\
\hline
\end{tabular}
\end{table*}
\section{INTRODUCTION}
Magnetic flux-transport dynamo modals have been successful for reproducing 
the many solar cycle features \citep[][and references therein]{dg06}.
The strength of the polar fields at the end
of a solar cycle  seems to be  an important
ingredient of   a kind of solar magnetic flux-transport dynamo modal and
using it as a `seed' in these modals
 the amplitude of Solar Cycle~24 was successfully predicted
\citep[e.g.][]{jiang07}. By using  
the strength of the polar fields at the end of a solar cycle
as  a  precursor for predicting the strength of
the next cycle  the  amplitudes of the last few cycles were 
successfully (with a reasonable uncertainty) predicted~\citep{pes08}.
  The amplitude of
the upcoming Solar Cycle~25 is also predicted by   a number of authors
 by simulating the strength of polar fields
at the end  of  Solar Cycle 24 and
most of these predictions
 indicate that Solar Cycle 25 will be similar strength as of
Solar Cycle~24~\citep[e.g.][]{cameron16,hath16,wang17,uh18,bn18}.
Recently, \cite{kumar21} used
the polar-field precursor method and predicted $126\pm3$ for the
 amplitude of Solar Cycle~25.

In a series of papers, \citep{jj07,jj08,jj15,jj21}, with   an 
  hypothesis that the transport of solar magnetic flux caused by solar
 rotational and meridional flows  may cause  the magnetic fields at a 
latitude during a time-interval of a solar cycle contribute to the magnetic 
fields at the same or a different latitude  during a time-interval  of the
 next solar cycle,  we   
determined the correlations between the sum of the areas of sunspot groups 
in different latitudes--and during different time intervals  of a
 solar cycle--and 
the amplitude of next solar cycle. 
This concept is somewhat close  to
 the concept of polar-field precursor method. 
We found  that the sum of the areas 
of sunspot groups in  $0^\circ-10^\circ$ latitude  interval  of the
southern hemisphere during a small interval  (7\,--\,9 months) just after
 one year from  the maximum of a solar cycle well-correlated to the
amplitude of the next solar cycle. This relationship was enabled us  to
predict the amplitudes of  Solar Cycles 24 and 25. The
exact physical reason behind this relationship is not clear yet,  but it  
 could be flux-transport dynamo mechanism. Therefore, the aforementioned
 sum of the areas of sunspot groups
 in a solar cycle  must have a
 relationship with the strength of polar fields at the end of the solar cycle
 (following minimum of the solar cycle).
 
There is usually more than one peak in a solar cycle.
Gnevyshev~(\citeyear{gne67}, \citeyear{gne77}) identified for the first
 time that the maximum  of  a  solar cycle contain two or more peaks
 and hence, they are known as Gnevyshev peaks. The level of solar activity
in the  time interval between  Gnevyshev peaks is known as the Gnevyshev gap
 \citep[see][]{sto03,ng10}.
 The level of solar activity in the Gnevyshev gap is relatively low and
this gap coincides with the   period of polarity of solar polar magnetic
reversal. Hence, it might be caused by the  global reorganization
of solar magnetic fields \citep{fs97,stori97}.
\citet{koz14} attributed  the cause of double
 maxima in solar cycles to the different behavior of large and small
 sunspot groups. According to \citet{baz00} the double or triple peaked 
maximum of a solar cycle may be due to the superposition of two
 quasi-oscillating 
processes with characteristic time-scales of 11 years and 1\,--\,3 years. 
\citet{du15} found that the double-peaked maxima of solar cycles may be caused
by a bi-dynamo mechanism. 
 \citet{phy17} have suggested  a cause of Gnevyshev gap may be due to
 spreading and  transfer of 
magnetic energy from higher to lower latitudes with progress of  solar cycle. 
 The presence of double peaks in the
smoothed time series of sunspot number or sunspot area could be caused by the
superposition of  slightly out of phase  northern and
southern hemispheres' sunspot indices. However, recent studies confirmed that
 the Gnevyshev gaps  occur in both the northern and the southern hemispheres'
 data and hence it is not an artifact of  superposition of out of phase
 sunspot indice of the hemispheres \citep{temm06,ng10,rj15,rcj21}. 
The double peak structure of the maximum of a solar cycle my have an 
implication on  geomagnetic activity \citep{ggt90}. Therefore, besides 
the amplitude (the value of main and highest peak),  predicting the 
 second maximum (the value of the second highest peak) of an upcoming solar
 cycle may be also important for better understanding the solar dynamo
 mechanism and the solar-terrestrial relationship. 
In the present analysis through hindsight we check 
 the consistency of the  above mentioned  relationship between the 
 sum of the areas of sunspot group in a solar cycle $n$  and the 
 amplitude of the next solar cycle ($n+1$). With the help of 
the predicted  amplitude of solar Cycle~25
 we   attempted to predict the value of 
 the second maximum  of Solar Cycle~25.

There exists a good correlation between the strength of the polar fields 
at the end of a solar cycle $n$ and amplitude of solar
 cycle $n+1$ \citep{sval05,jiang07}. There also exists  a good-correlation 
 between the aforementioned 
sum of the area of sunspot groups in the solar cycle $n$ and 
the amplitude of the solar cycle $n+1$.  
  Hence, one can expect the existence of a good correlation 
between the strength of polar fields at the end of a solar cycle 
 and the aforementioned sum of the areas of  sunspot groups in 
the solar cycle. Using 
the latter as a precursor  it is possible to predict the amplitude of
 a solar cycle much 
earlier (by 3\,--\,4 years) than that  by using the former.
In addition, the latter may also have a power of
 prediction of  the  strength of the polar fields at the end of the 
solar cycle by 3\,--\,4 years in advance. In the present analysis our aim
 is also to investigate whether this is possible or not,
and to find a plausible reason behind the difference between 
the predicted values of the  amplitude of Solar Cycle 25 made by 
using these two different  precursors. 

In the next section we describe the data and analysis. 
In Sec.~3 we describe the results, and in Sec.4 we present 
the conclusions and discuss them briefly.

\begin{figure}
\centering
\includegraphics[width=7.9cm]{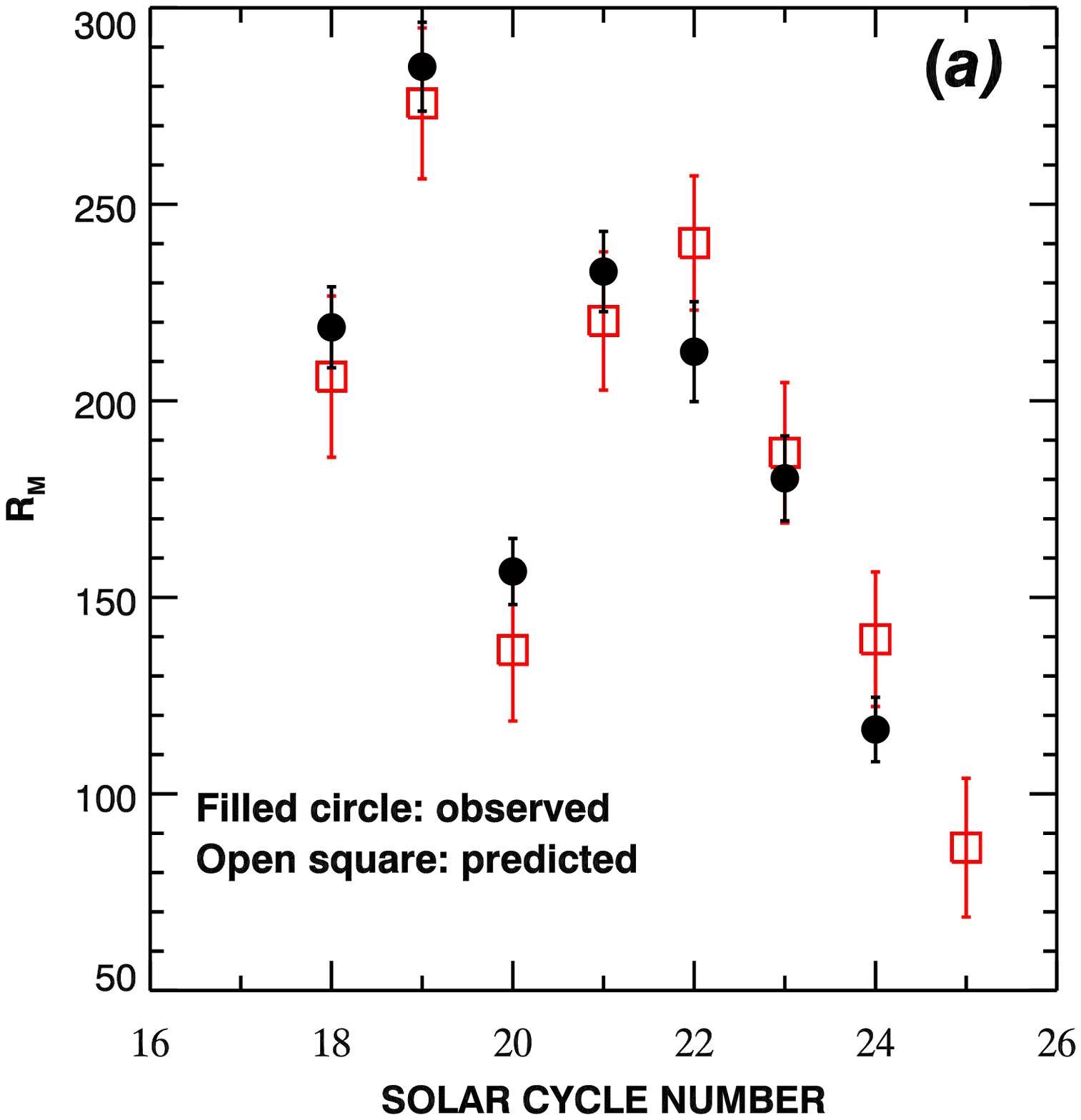}
\includegraphics[width=7.9cm]{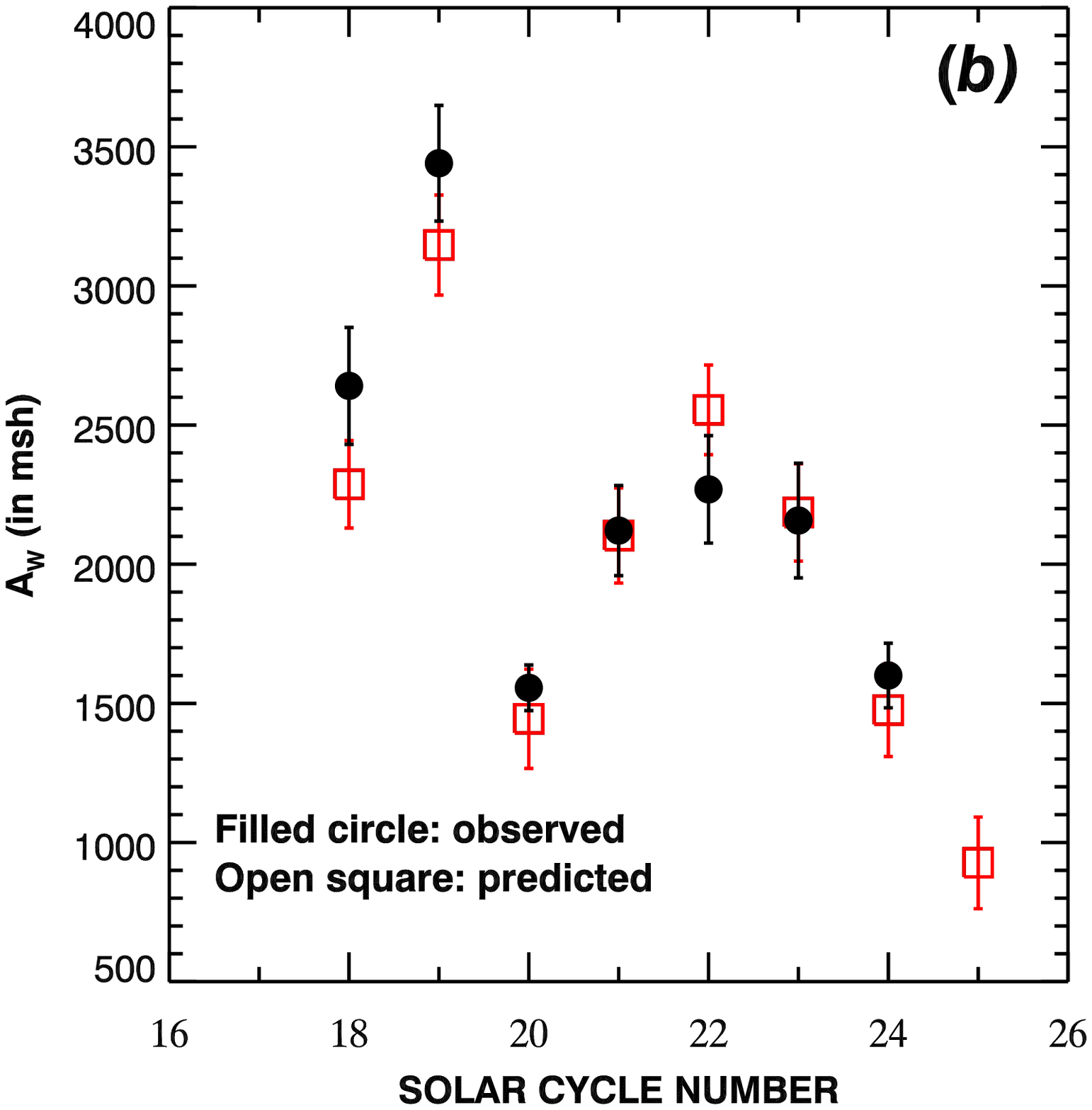}
\caption{Hindsight: Comparison of the observed  and the 
predicted values  ({\bf a}) of $R_{\rm M}$  and ({\bf b}) of  
 $A_{\rm W}$  of Solar Cycles 18--24. The predicted 
values of $R_{\rm M}$ and $A_{\rm W}$ of Solar Cycle~25 
are also shown.} 
\label{f1}
\end{figure}

\begin{figure}
\centering
\includegraphics[width=8cm]{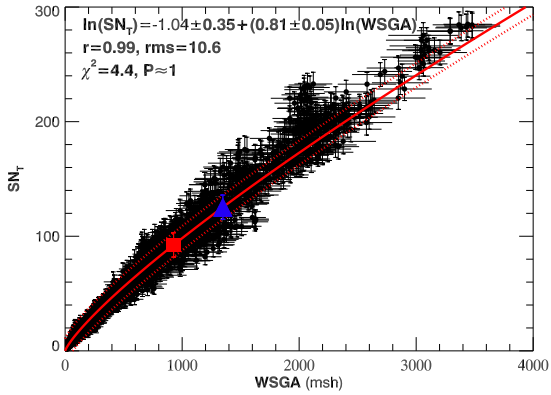}
\caption{Scatter plot of the 13-month smoothed monthly mean area of sunspot 
groups in the Sun's whole sphere
(WSGA) versus the  13-month smoothed monthly mean  
 ${\rm SN}_{\rm T}$ during the period 1874\,--\,2017
(1713 data points). The  continuous curve (red) represents
 the linear least-squares best-fit to the $\ln ({\rm WSGA})$ and
 $\ln ({\rm SN}_{\rm T})$.
 The dotted curve (red) represents the one-rms  
 level.
 The obtained linear equation and
the values of the corresponding correlation coefficient $r$, rms, and
$\chi^2$, and $P$ are  given. The
filled-triangle  (blue) and filled-square (red) 
  represent the predicted value of
${\rm SN}_{\rm T}$, i.e. $R_{\rm M}$  of Solar Cycle~25, 
by using the  values of 
$A_{\rm W}$   of Solar Cycle~25 that are predicted using the 
 $A_{\rm W}$--$W_{\rm M}$ relation in Javaraiah (2022)  and 
$A^*_{\rm W}$--$W_{\rm W}$ relation above (cf. Table~2),
 respectively.}
\label{f2}
\end{figure}

\section{DATA AND ANALYSIS}
Here have used  monthly and  13-month smoothed monthly mean
 Version-2 international sunspot number (SN)  during the period 
October 1874\,--\,June 2017 (we downloaded  the files
  SN\_m\_tot\_v2.0.txt and  
 SN\_ms\_tot\_v2.0.txt from {\sf www.sidc.be/silso/datafiles}).
The details of changes and corrections in Version-2 SN can be found 
 in \cite{clette16}.
 We have used the values of the
 amplitudes ($R_{\rm M}$), i.e.
 the highest values of 13-month smoothed monthly mean sunspot numbers, and
the maximum epochs ($T_{\rm M}$) of Sunspot Cycles~12\,--\,24 given
 by \cite{pesnell18}.
 \cite{pesnell18} determined these 
 from the time series of  13-month smoothed monthly mean
values of SN.
From the same  time series  
 we determined  the epoch ($T_{\rm S}$) and the value of second largest 
peak ($S_{\rm M}$, say) during the maximum phase of each of
 Sunspot Cycles~12\,--\,24.

 Recently, \citep{jj21}, we analysed the daily sunspot-group  data  reported by the  Greenwich
 Photoheliographic Results (GPR) during the period 1874\,--\,1976,
 Debrecen Photoheligraphic Data (DPD) during the period 1977\,--\,2017,
and the revised Version-2 SN during
the period 1874\,--\,2017. We determined
 the correlation of  $R_{\rm M}$, $i.e.$ 
the amplitudes  of Solar Cycles 13\,--\,24, 
with  the  sum of the areas of the
sunspot groups  in different  $10^\circ$ latitude intervals and in
different time intervals during  
Solar Cycles 12\,--\,23.  We found that the sum of the areas 
($A^*_{\rm R}$)
of sunspot groups in $0^\circ - 10^\circ$ latitude interval of the southern 
hemisphere during a small  (7-month) interval just after  
one year from the maximum epoch of a solar cycle $n$ has
 a maximum correlation
with $R_{\rm M}$ of the next solar cycle $n+1$.  We derived 
 the linear relationship between $A^*_{\rm R} (n)$ and $R_{\rm M} (n+1)$ 
 by the method of   linear least-square fit.  By using the obtained 
$A^*_{\rm R} (n) - R_{\rm M} (n+1)$ linear relationship and $A^*_{\rm R}$ of   
 Solar Cycle 24, we predicted the value $86 \pm 18$ for 
 $R_{\rm M}$ of  Solar Cycle 25. 
Similarly, a prediction was also made for 
 $A_{\rm W}$, i.e. the 13-month smoothed monthly
mean areas of sunspot groups at  $T_{\rm M}$  of a Solar Cycle~25.
Here  we  check the consistency of the aforementioned method 
through  hindsight of the  $A^*_{\rm R} (n)  - R_{\rm M} (n+1)$ 
relationship and also the $A^*_{\rm W} (n)- A_{\rm W} (n+1)$ relationship, 
where $A^*_{\rm W} (n)$ is the sum of the areas of the sunspot  groups, 
  determined similarly  as $ A^*_{\rm R} (n)$,  well correlated  with 
 $A_{\rm W} (n+1)$.     

\begin{figure*}
\centering
\includegraphics[width=\textwidth]{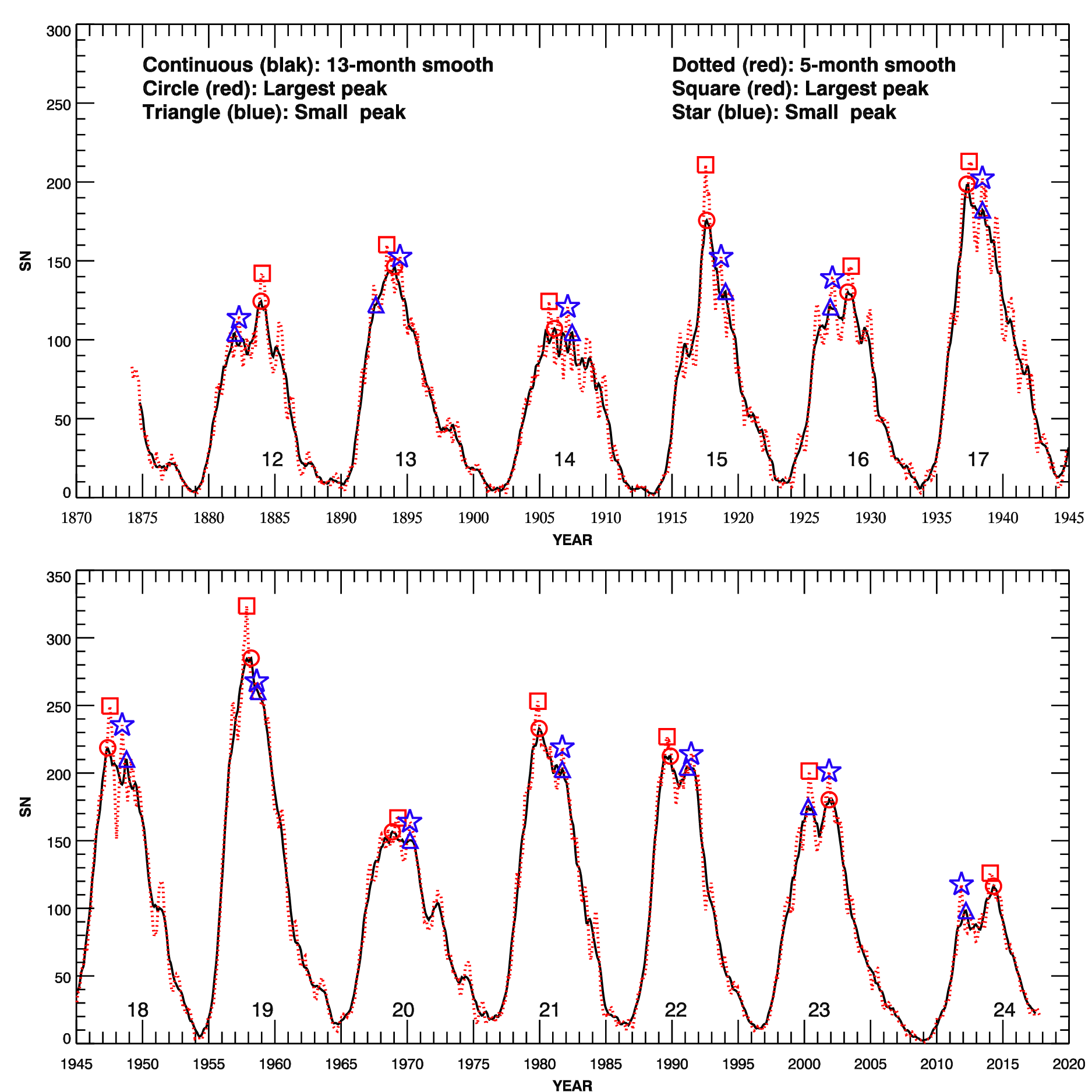}
\caption{Variations in the   5-month (dotted-curve) and 
13-month (continuous curve) smoothed monthly mean sunspot number 
(SN) during the period 1874\,--\,2017. 
The symbols {\it circle} ({\it red})
 and {\it triangle} ({\it blue}) 
represent the largest and the second largest peaks of a sunspot cycle
 in the 13-month smoothed series. The corresponding peaks in 
5-month smoothed series are represented by  
the symbols {\it square} ({\it red})
 and {\it star} ({\it blue}), respectively.  
The Waldmeier  solar cycle number is also given.} 
\label{f3}
\end{figure*}

We find the
 existence of a high correlation and a good linear 
relationship  between the cycle-to-cycle 
modulations in $R_{\rm M}$  and $S_{\rm M}$.
By using this 
relation and the values predicted for $R_{\rm M}$ of Solar Cycle~25 by 
\citet{jj21,jj22} we predict the value of 
$S_{\rm M}$ of  Solar Cycle~25. 
 In \citet{jj22} we have calculated the least-square cosine fits to the
 cycle-to-cycle modulation in $R_{\rm M}$ during Solar Cycles 12\,--\,24.
  The same calculations are done here for $S_{\rm M}$. 
 Since there is ambiguity  in  the positions of  $S_{\rm M}$ of 
some cycles  determined 
from the  13-month smoothed monthly mean SN series, hence we also 
determined 5-month smoothed monthly mean SN series and  using it
 repeated all the calculations.
In order to find that whether the peak of $R_{\rm M}$ or 
that of $S_{\rm M}$ would be 
first during the maximum  of Solar Cycle~25, we fit cosine curves to the 
values of peaks that occurred first and second during the maxima 
  of Solar Cycles~12\,--\,24.

Although it is well believed that the strength of polar magnetic fields at 
the end  of a solar cycle  is a good precursor for predicting the 
amplitude of the next solar cycle~\citep{sch78,sval05}, it is not  clear yet
 exactly the  time 
of polar fields which predict the amplitude. Therefore, 
the predicted  amplitude of solar cycle has a considerable large
 uncertainty~\citep{sval05}.
\cite{sval05} analysed the  polar-fields data measured  in Wilcox 
Observatory (WCO) and  Mt. Wilson Observatory (MWO) during 1970\,--\,2005.
They have used 
the average strength of dipole moment (DM: the
average unsigned difference between the north and south polar fields)  
in the three years before the end of each of Solar Cycles 20\,--\,23 
(one year in the case of Solar Cycle 23) for predicting the 
amplitude ($R_{\rm M}$) of Solar Cycle 24.
Here we have analysed the  polar-fields data  measured in WCO and besides 
determining the average values of  DM  of the three years before the 
 end of each of  Solar Cycles 20\,--\,23, the average value of DM of the 
three years before  the end of Solar Cycle 24 is determined. We have used 
the value of DM around the end, December/2019, of Solar Cycle~24. 
 The WCO data are available at {\tt wso.stanford.edu/Polar.html} are
 30-day averages of the
 magnetic field measured in the polemost aperture calculated every 10 days.
  We have used the data that are 
corrected for the Earth's rotational frequency.  We  have taken the 
corresponding  average value of DM of Solar Cycle~20 from 
Table~1 in \cite{jiang07}, it was determined from MWO data by \cite{sval05}.
We determined  correlation and linear least-square-fit
to the values of DM and $A^*_{\rm R}$  of  Solar  
Cycles 20\,--\,23. By using the  obtained linear relationship 
 first we predicted the average value of DM of the 
three years before the end of Solar cycle~24. We determined the 
correlation and the 
linear least-square fit of  ${\rm DM} (n)$ and $R_{\rm M} (n+1)$, by using 
the values of DM of  Solar  Cycles 20\,--\,23 and the values of 
$R_{\rm M}$  of Solar Cycles~21\,--\,24. By
substituting in the  ${\rm DM} (n) - R_{\rm M} (n+1)$ relation  
 the predicted and observed values of DM of 
Solar Cycle~24, we obtained the corresponding values for $R_{\rm M}$ of 
Solar Cycle~25. Finally we check the correlation between  DM 
and $A^*_{\rm R}$ values of  all five solar cycles.

\begin{table*}
\caption{The epochs  $T_{\rm M}$ and $T_{\rm S}$  of  
  $R_{\rm M}$ and  $S_{\rm M}$, respectively,  of Sunspot Cycles 12\,--\,24 
 determined from 13-month smoothed monthly mean SN.
 The intervals (Gnevyshev gaps, in year) between these
  peaks, the ratios $S_{\rm M}$/$R_{\rm M}$, and the values of the
 mean and standard deviation of the
 absolute values of these parameters are also given.} 
\label{table3}
\begin{tabular}{lcccccccc}
\hline
$n$ &$T_{\rm M}$&$R_{\rm M}$&$\sigma_{\rm M}$&$T_{\rm S}$ &$S_{\rm M}$ 
&$\sigma_{\rm S}$&$T_{\rm M}-T_{\rm S}$& $S_{\rm M}$/$R_{\rm M}$\\
\hline
 12 &1883.96 &124.4& 12.5& 1881.96& 104.1&  11.5& $-$2.00& 0.84\\
 13 &1894.04 &146.5& 10.8& 1892.62& 122.2&  12.1& $-$1.42& 0.83\\
 14 &1906.12 &107.1&  9.2& 1907.45& 104.6&   9.1&   1.33& 0.98\\
 15 &1917.62 &175.7& 11.8& 1919.04& 130.6&  10.2&   1.42& 0.74\\
 16 &1928.29 &130.2& 10.2& 1926.96& 120.8&   9.8& $-$1.33& 0.93\\
 17 &1937.29 &198.6& 12.6& 1938.45& 182.3&  12.0&   1.17& 0.92\\
 18 &1947.37 &218.7& 10.3& 1948.79& 210.3&   9.7&   1.42& 0.96\\
 19 &1958.20 &285.0& 11.3& 1958.71& 260.3&  10.8&   0.50& 0.91\\
 20 &1968.87 &156.6&  8.4& 1970.20& 150.3&   8.2&   1.33& 0.96\\
 21 &1979.96 &232.9& 10.2& 1981.71& 202.7&  13.3&   1.75& 0.87\\
 22 &1989.87 &212.5& 12.7& 1991.12& 204.4&  12.5&   1.25& 0.96\\
 23 &2001.87 &180.3& 10.8& 2000.29& 175.2&  10.5& $-$1.58& 0.97\\
 24 &2014.29 &116.4&  8.2& 2012.21&  98.3&   7.5& $-$2.08& 0.84\\
\hline
Mean&&    175.8& 52.5&& 158.9&  50.8&1.43$\pm$0.39 &0.90$\pm$0.07\\
\hline
\end{tabular}
\end{table*}

\begin{figure}
\centering
\includegraphics[width=8.5cm]{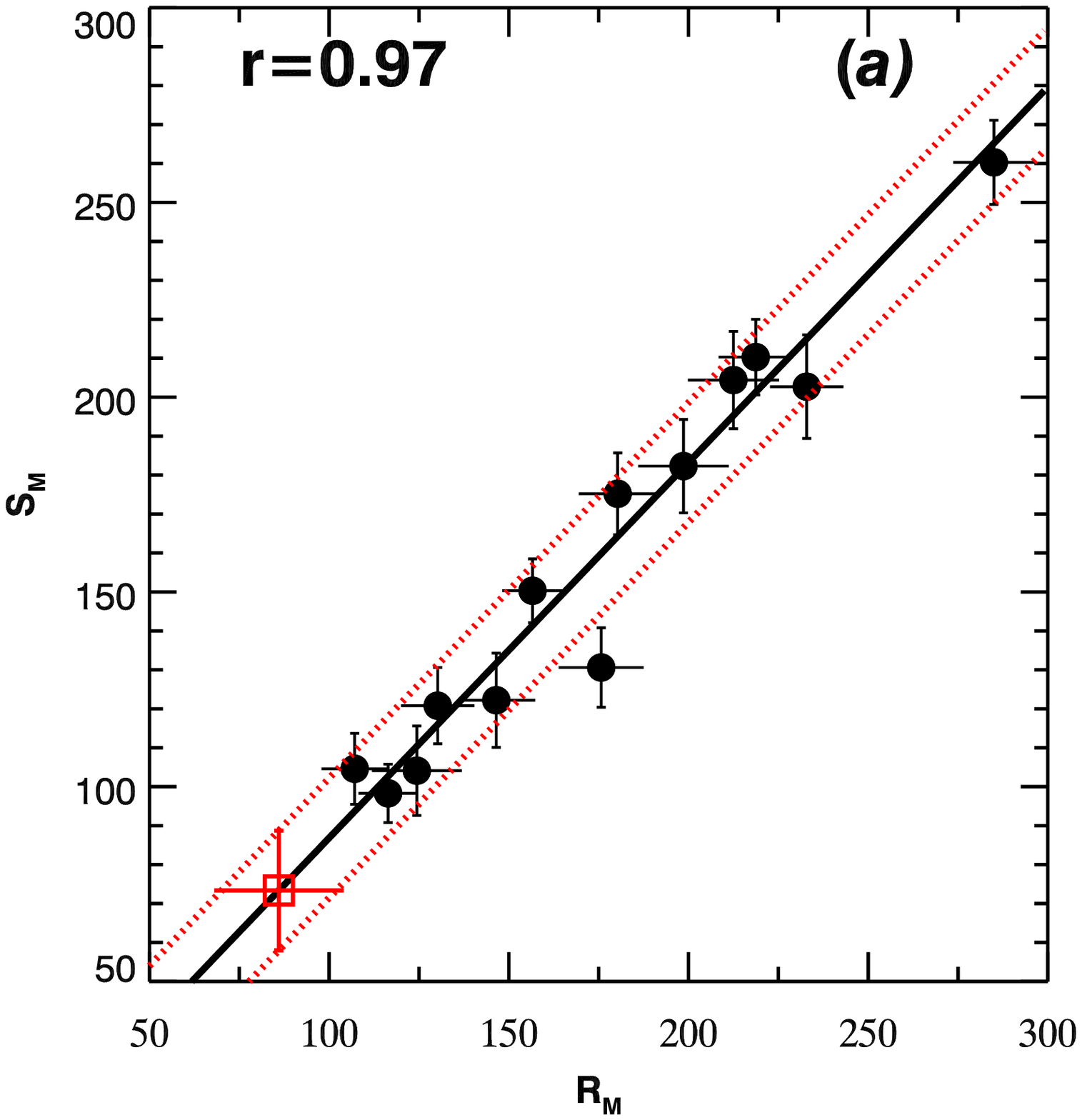}
\includegraphics[width=8.5cm]{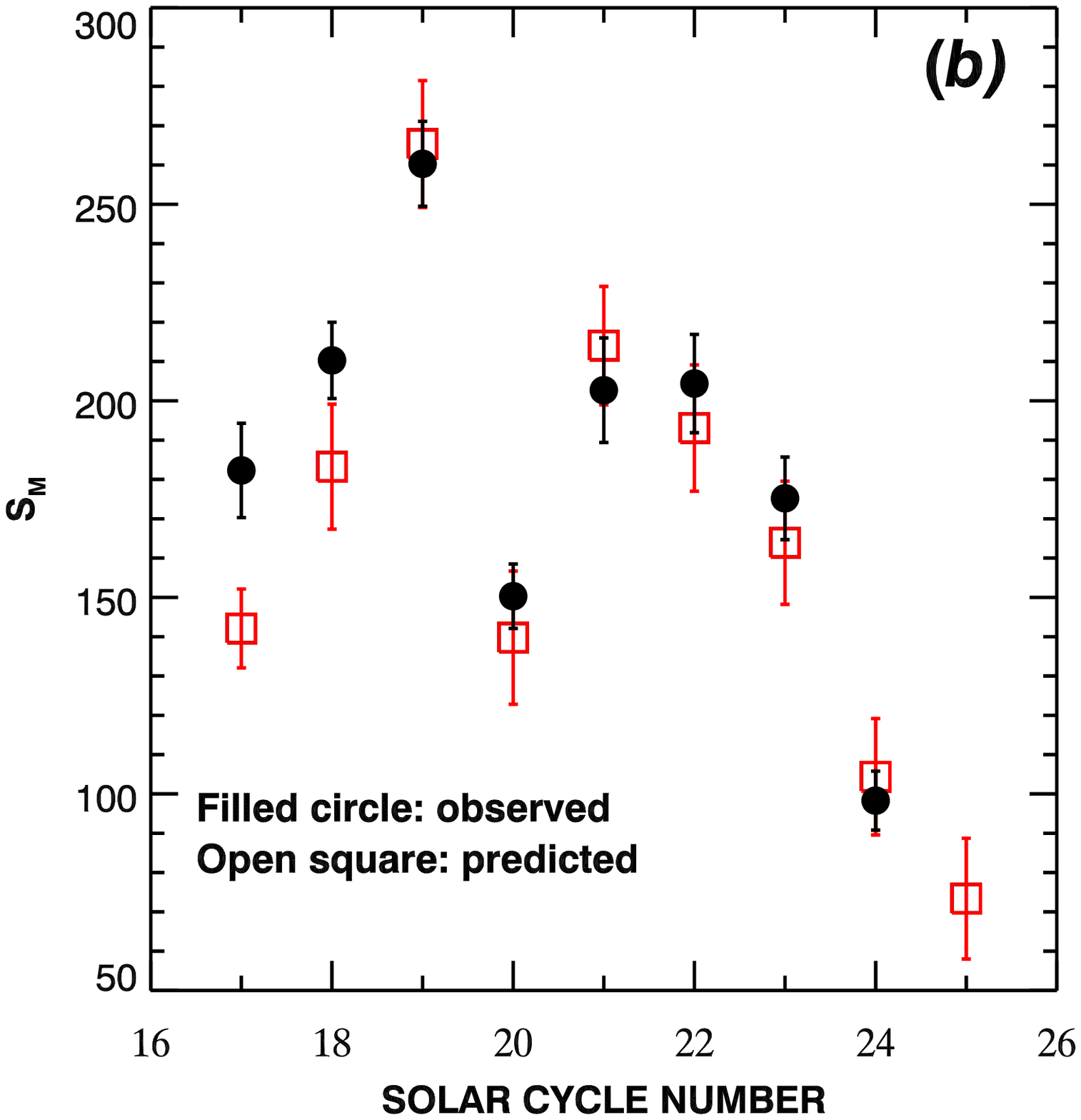}
\caption{({\bf a}) Correlation between $R_{\rm M}$ and $S_{\rm M}$
(the values given Table~\ref{table3}) during
Solar Cycles~12\,--\,24. The {\it continuous line} represents 
the best-fit linear relationship, Equation~(1). 
The {\it dotted lines} ({\it red}) are drawn at one-rms level. 
   ({\bf b}) Hindsight: comparison of the observed 
and the predicted values  of $S_{\rm M}$. The predicted
value ({\it red square}) of $S_{\rm M}$  of Solar Cycle~25 is 
also shown in both ({\bf a}) and ({\bf b}) .}
\label{f4}
\end{figure}

\begin{table*}
\caption[]{Hindsight: The values of intercept ($C$)
and slope ($D$) of the linear
relationship between $R_{\rm M}$  and
 $S_{\rm M}$ correspond to the predictions for $S_{\rm M}$ of
  Solar Cycles 17\,--\,25. In the 
case of Solar Cycle $n = 25$ the predicted value of $R_{\rm M}$ 
is used. The corresponding values of the correlation coefficient ($r$),
$\chi^2$and its probability ($P$), number of data points ($N$),
and predicted values of $S_{\rm M}$ are also given.}
\label{table4}
\begin{tabular}{lccccccccc}
\hline
  \noalign{\smallskip}
$n$ & $C$ & $D$& $r$ & $\chi^2$& $P$&  $N$ & Pred.$S_{\rm M}$ \\
\hline
  \noalign{\smallskip}
 17&$ 61.00\pm 30.46$&$ 0.41\pm0.22$&  0.89& 0.81& 0.85&  5&$ 142.1\pm10.0$\\
 18&$ 14.23\pm 28.36$&$ 0.77\pm0.19$&  0.91& 3.86& 0.42&  6&$ 183.2\pm15.9$\\
 19&$-12.75\pm 23.64$&$ 0.98\pm0.15$&  0.94& 5.37& 0.37&  7&$ 265.3\pm16.2$\\
 20&$ -9.62\pm 17.07$&$ 0.95\pm0.09$&  0.97& 5.41& 0.49&  8&$ 139.8\pm16.9$\\
 21&$ -6.97\pm 16.68$&$ 0.95\pm0.09$&  0.97& 6.11& 0.53&  9&$ 214.0\pm15.1$\\
 22&$ -4.67\pm 16.05$&$ 0.93\pm0.09$&  0.97& 6.50& 0.59& 10&$ 193.1\pm16.1$\\
 23&$ -5.98\pm 16.09$&$ 0.94\pm0.09$&  0.97& 6.89& 0.65& 11&$ 163.9\pm15.7$\\
 24&$ -5.79\pm 16.08$&$ 0.95\pm0.09$&  0.97& 7.43& 0.68& 12&$ 104.4\pm14.8$\\
 25&$ -9.54\pm 14.58$&$ 0.96\pm0.08$&  0.97& 7.66& 0.74& 13&$  73.4\pm15.4$\\
\hline
\end{tabular}
\end{table*}

\begin{figure}
\centering
\includegraphics[width=8.5cm]{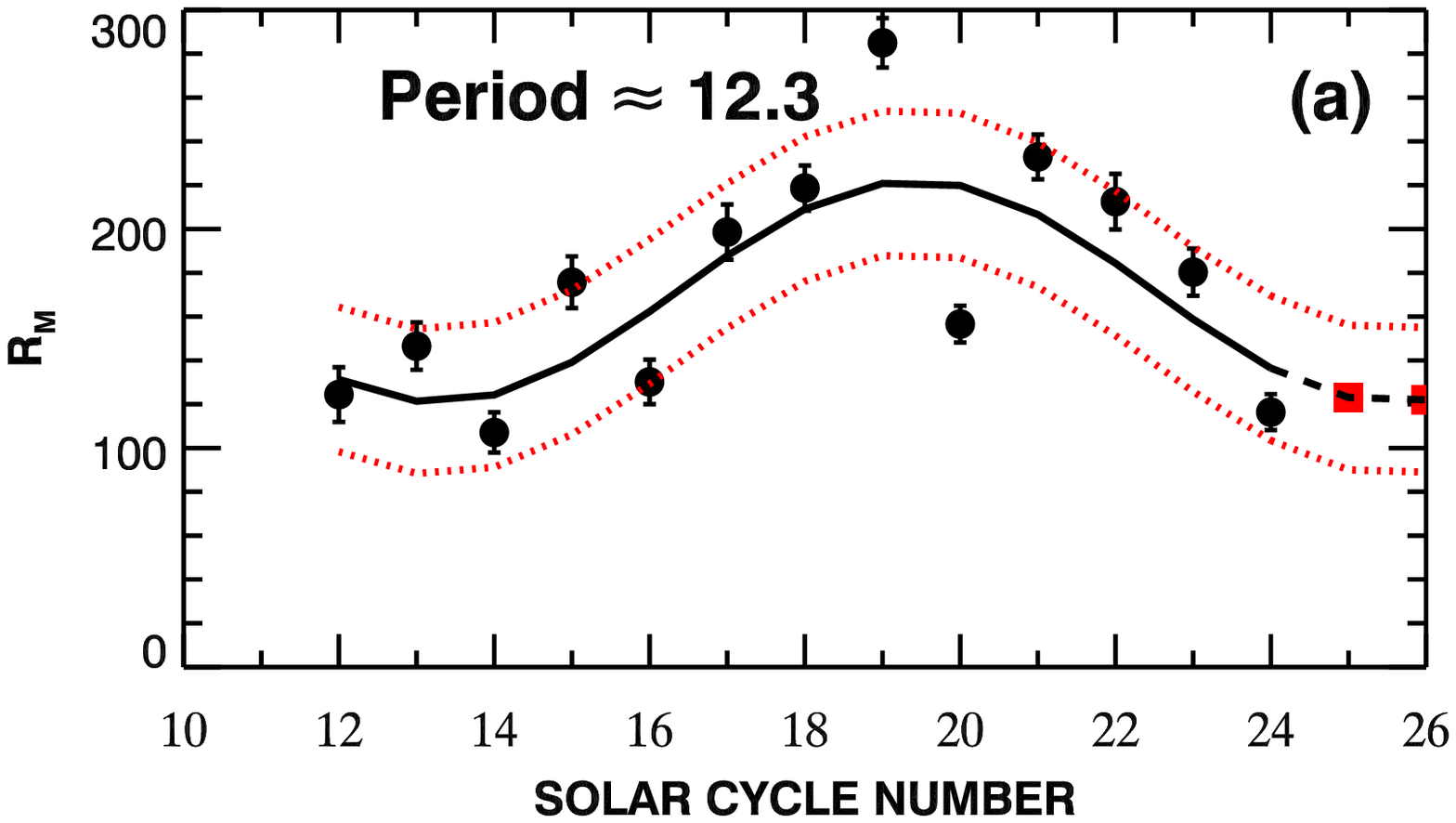}
\includegraphics[width=8.5cm]{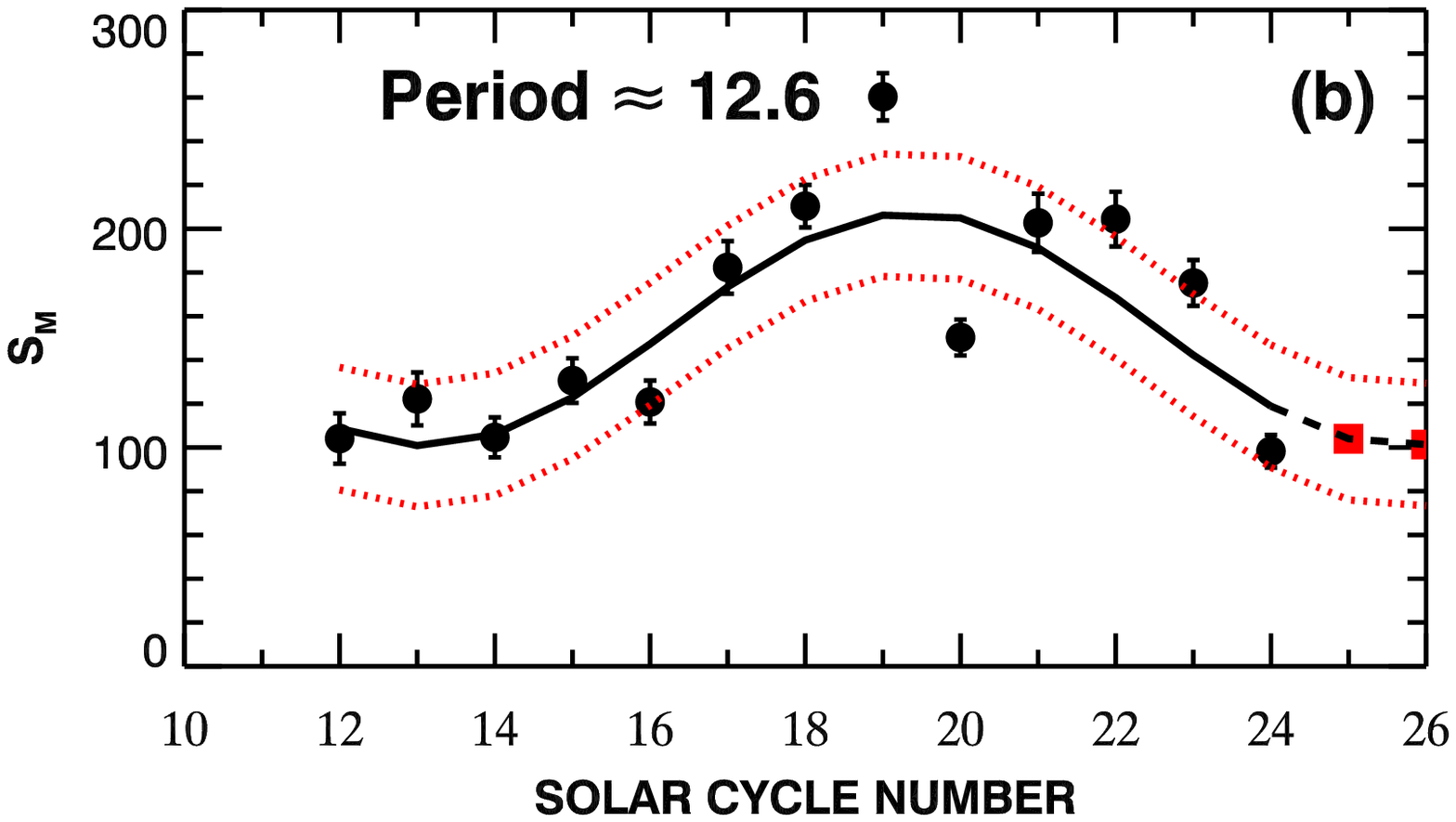}
\caption{{\it Continuous curve} represents the best-fit cosine function
to the values  ({\it filled circles})  ({\bf a}) of
 $R_{\rm M}$ and ({\bf b}) of $S_{\rm M}$   of  Sunspot Cycles 12\,--\,24 
(for values in Table~\ref{table3}).
The  {\it dotted curve} ({\it red})
represents the one-rms  level. The  extrapolated portion is shown as a
{\it dashed curve} and  the {\it filled squares} ({\it red}) represent
 the predicted values  of $R_{\rm M}$ and $S_{\rm M}$     of 
Sunspot Cycles~25 and 26. The  period (in number of solar cycles)  of 
the cosine function is also shown.}
\label{f5}
\end{figure}

\begin{figure}
\centering
\includegraphics[width=8.5cm]{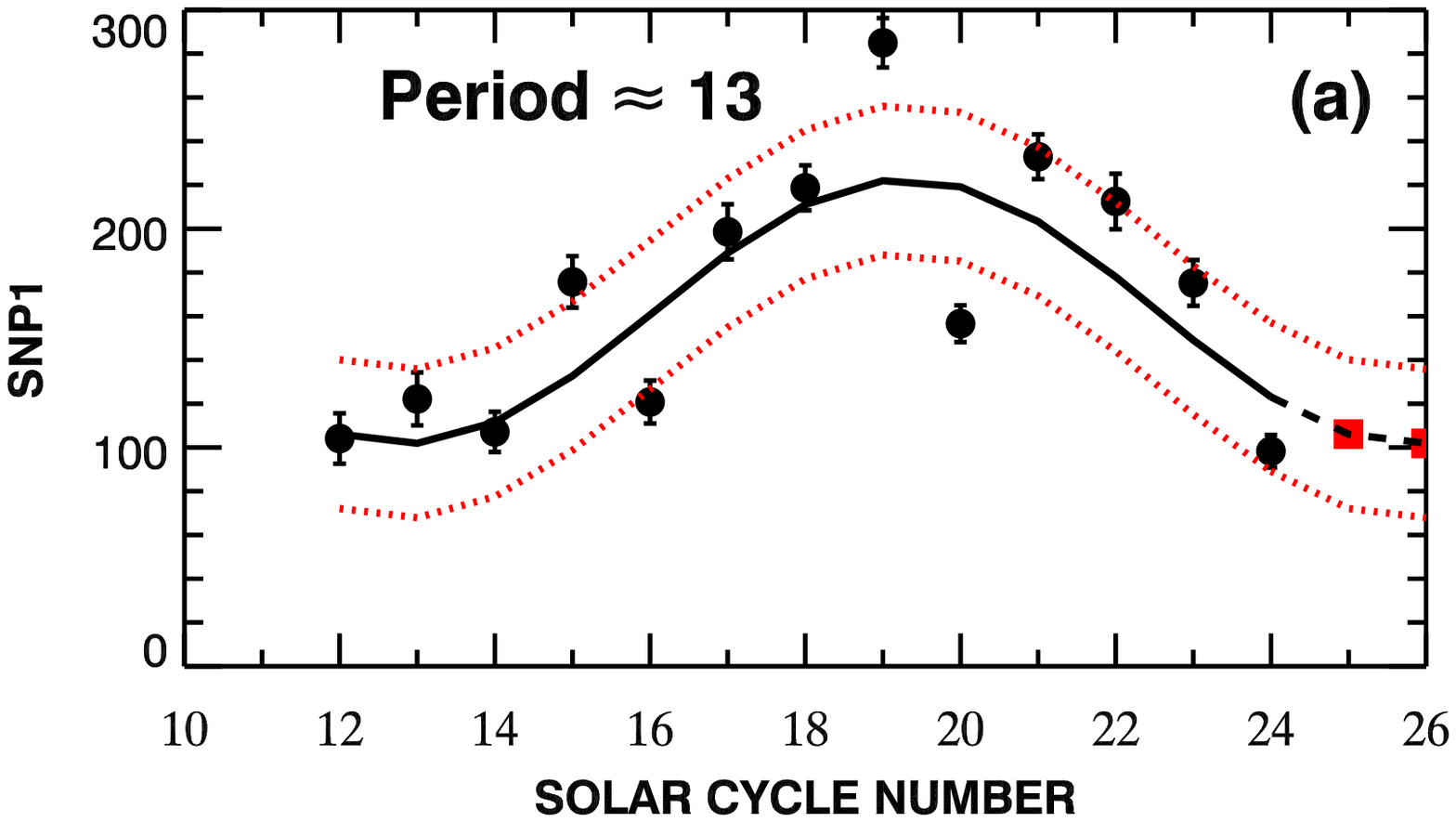}
\includegraphics[width=8.5cm]{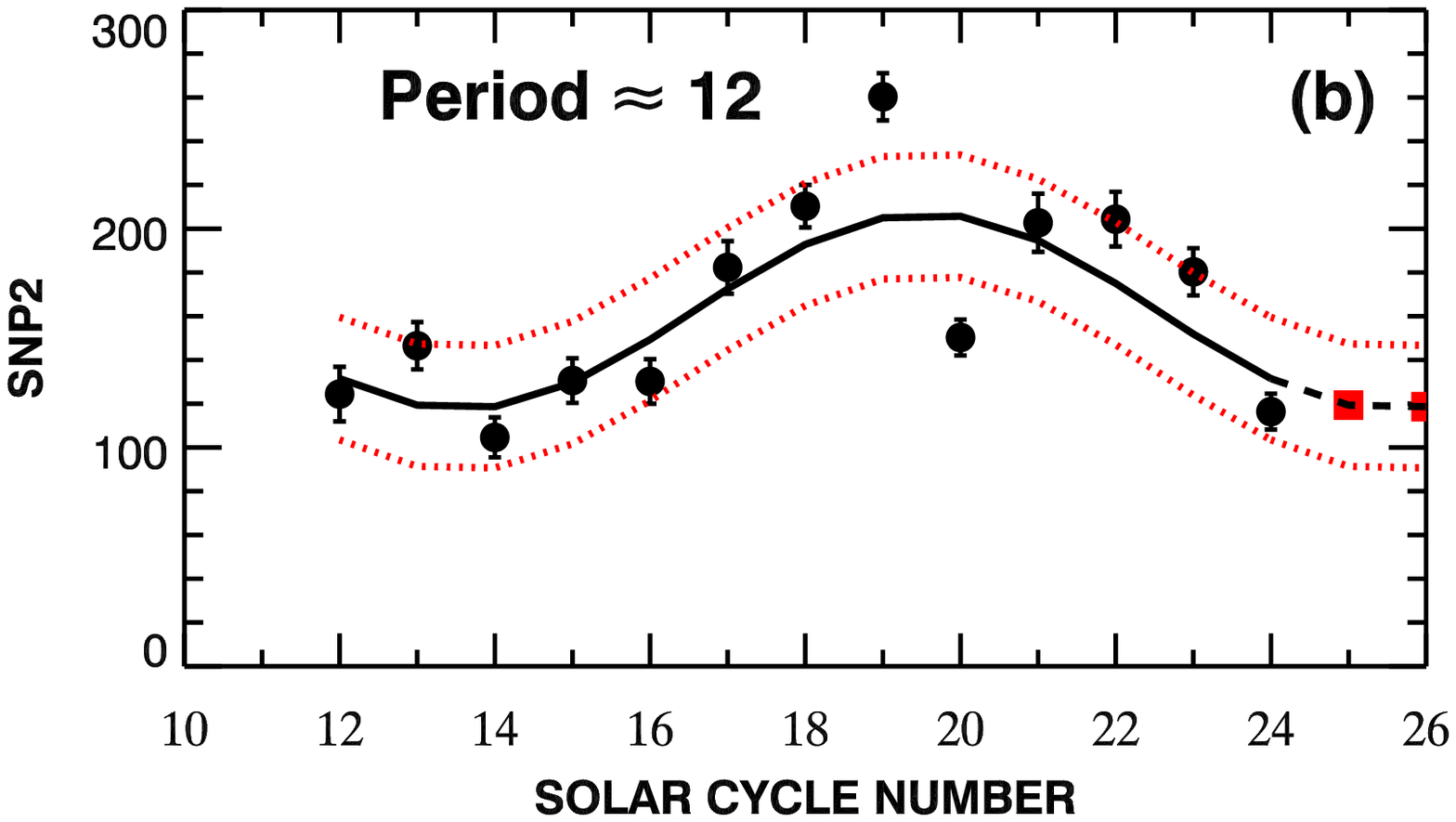}
\caption{{\it Continuous curve} represents the best-fit cosine function
to the values  ({\it filled circles})  ({\bf a}) of
 SNP1 and ({\bf b}) of SNP2   of  Sunspot Cycles 12\,--\,24 
(for values in Table~\ref{table5}).
The  {\it dotted curve} ({\it red})
represents the one-rms  level. The  extrapolated portion is shown as a
{\it dashed curve} and  the {\it filled squares} ({\it red}) represent
 the predicted values  of SNP1 and SNP2   of 
Sunspot Cycles~25 and 26. The  period (in number of solar cycles)  of 
the cosine function is also shown. 
}
\label{f6}
\end{figure}

\begin{figure}
\centering
\includegraphics[width=8.5cm]{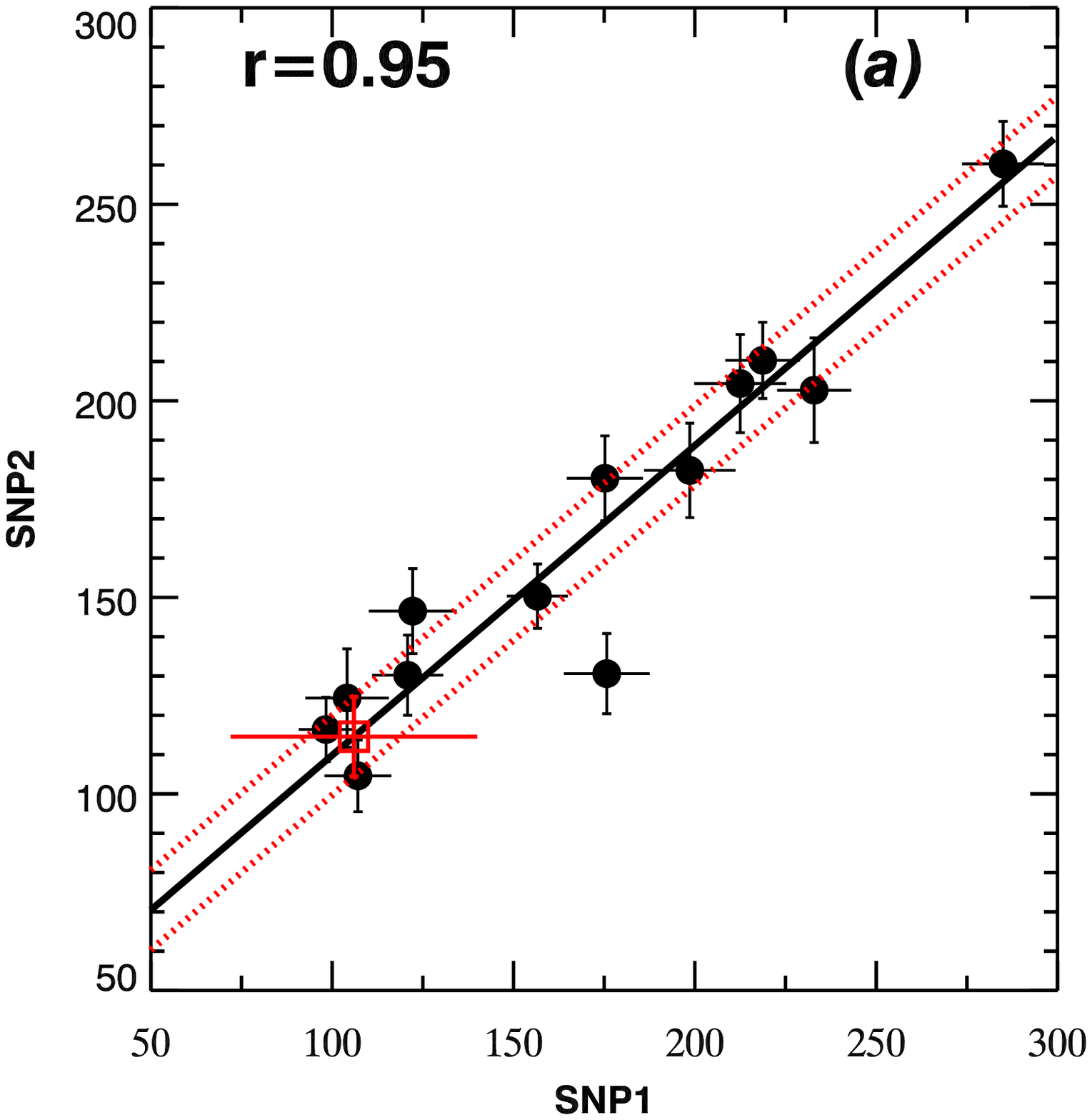}
\includegraphics[width=8.5cm]{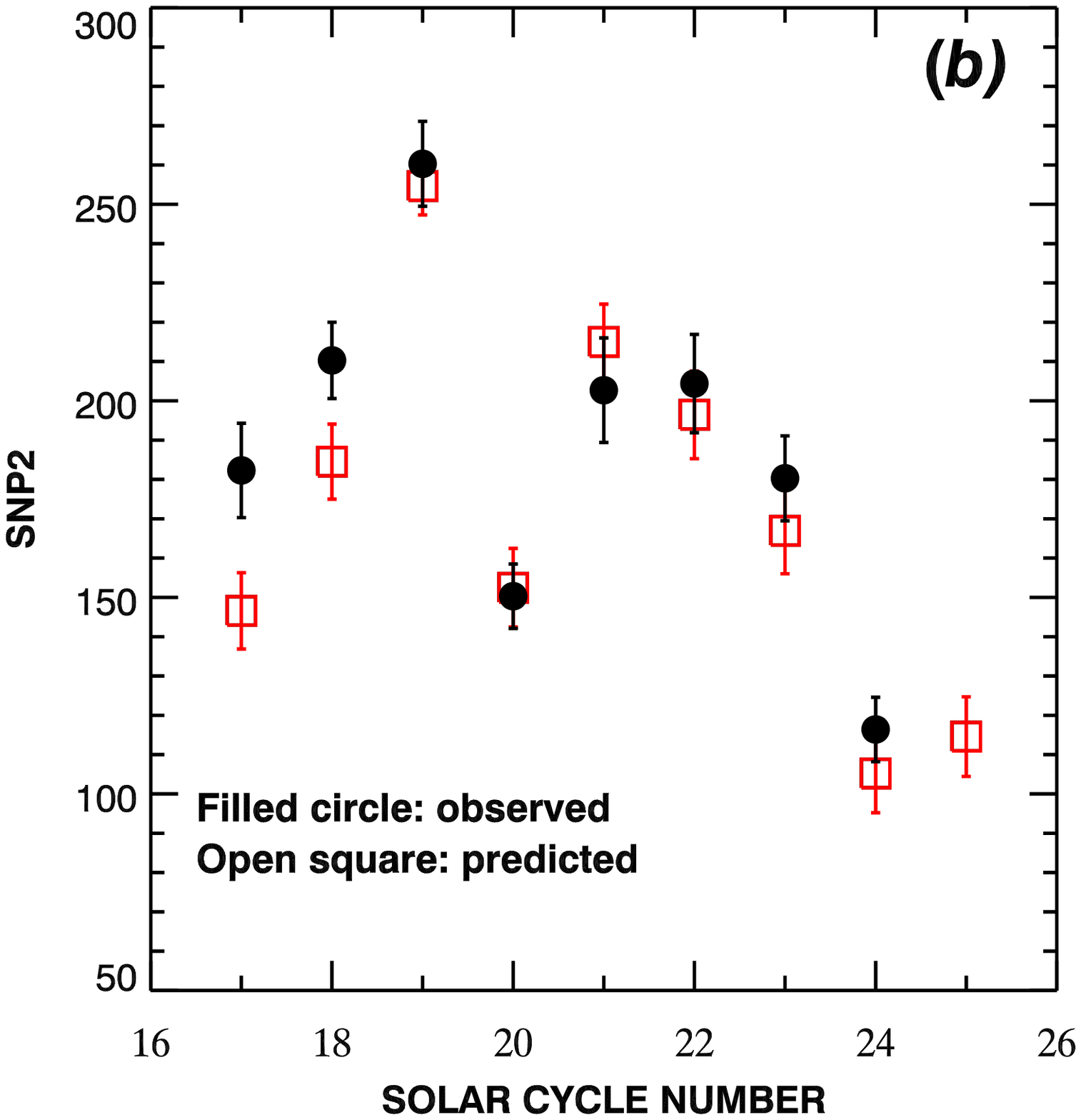}
\caption{({\bf a}) Correlation between SNP1 and SNP2 during
Solar Cycles~12\,--\,24. The {\it continuous line} represents
the best-fit linear relationship, Equation~(2).
The {\it dotted lines} ({\it red}) are drawn at one-rms level. ({\bf b}) Hindsight: comparison of the observed and the
predicted values  of SNP2. The predicted
value of SNP2 of Solar Cycle~25 is also shown in both ({\bf a}) and 
({\bf b}).}
\label{f7}
\end{figure}

\begin{table*}
\caption{The epochs  TSN1 and TSN2 of the first peak (SNP1) and the second 
peak (SNP2), respectively, of Sunspot Cycles 12\,--\,24
(determined from 13-month smoothed monthly mean SN).
 The intervals (Gnevyshev gaps, in year) between these
  peaks, the ratios SNP1/SNP2, and the values of the corresponding 
 mean and standard deviation are also given.
The values of $R_{\rm M}$  are indicated with bold-font.}
\label{table5}
\begin{tabular}{lcccccccc}
\hline
$n$ &TSN1&SNP1&$\sigma_1$&TSN2&SNP2&$\sigma_2$&TSN2$-$TSN1& $\frac{\rm SNP1}{\rm SNP2}$\\
\hline
 12&   1881.96&104.1& 11.5&1883.96&{\bf 124.4}& {\bf 12.5}&2.00&0.84\\
 13&   1892.62&122.2& 12.1&1894.04&{\bf 146.5}& {\bf 10.8}&1.42&0.83\\
 14&   1906.12&{\bf 107.1}& {\bf 9.2}&1907.45&104.6&  9.1&1.33&1.02\\
 15&   1917.62&{\bf 175.7}& {\bf 11.8}&1919.04&130.6& 10.2&1.42&1.35\\
 16&   1926.96&120.8&  9.8&1928.29&{\bf 130.2}& {\bf 10.2}&1.33&0.93\\
 17&   1937.29&{\bf 198.6}& {\bf 12.6}&1938.45&182.3& 12.0&1.17&1.09\\
 18&   1947.37&{\bf 218.7}& {\bf 10.3}&1948.79&210.3&  9.7&1.42&1.04\\
 19&   1958.20&{\bf 285.0}& {\bf 11.3}&1958.71&260.3& 10.8&0.50&1.09\\
 20&   1968.87&{\bf 156.6}& {\bf 8.4}&1970.20&150.3&  8.2&1.33&1.04\\
 21&   1979.96&{\bf 232.9}& {\bf 10.2}&1981.71&202.7& 13.3&1.75&1.15\\
 22&   1989.87&{\bf 212.5}& {\bf 12.7}&1991.12&204.4& 12.5&1.25&1.04\\
 23&   2000.29&175.2& 10.5&2001.87&{\bf 180.3}& {\bf 10.8}&1.58&0.97\\
 24&   2012.21&98.3&  7.5&2014.29&{\bf 116.4}&  {\bf 8.2}&2.08&0.84\\
\hline
Mean&&    169.8&58.1&& 164.9& 45.9&1.43$\pm$0.39 &1.02$\pm$0.14\\
\hline
\end{tabular}
\end{table*}

\begin{table*}
\caption[]{Hindsight: The values of intercept ($C$)
and slope ($D$) of the linear
relationship between SNP1  and
 SNP2 correspond to the predictions for SNP2 of
  Solar Cycles 17\,--\,25. In the 
case of Solar Cycle $n = 25$ the value of SNP1 predicted 
through the extrapolation of best fit cosine curve of SNP2 shown in 
Fig.~\ref{f6}(b) is used. The corresponding values of the correlation 
coefficient ($r$), $\chi^2$ and its probability ($P$), number of 
data points ($N$), and predicted values of SNP2 are also given.}
\label{table6}
\begin{tabular}{lccccccccc}
\hline
  \noalign{\smallskip}
$n$ & $C$ & $D$& $r$ & $\chi^2$& $P$&  $N$ & Pred. SNP2 \\
\hline
  \noalign{\smallskip}
 17&$ 89.32\pm 28.94$&$ 0.29\pm 0.23$& 0.33&  7.69& 0.05& 5&$146.6\pm 9.7$\\
 18&$ 49.45\pm 23.99$&$ 0.62\pm 0.17$& 0.76&  9.84& 0.04& 6&$184.5\pm 9.5$\\
 19&$ 26.73\pm 19.14$&$ 0.80\pm 0.12$& 0.87& 11.36& 0.04& 7&$254.6\pm 7.3$\\
 20&$ 23.53\pm 14.57$&$ 0.82\pm 0.08$& 0.94& 11.42& 0.08& 8&$152.5\pm10.0$\\
 21&$ 23.01\pm 14.34$&$ 0.82\pm 0.08$& 0.94& 11.46& 0.12& 9&$215.0\pm 9.6$\\
 22&$ 25.18\pm 13.87$&$ 0.81\pm 0.08$& 0.94& 11.96& 0.15&10&$196.4\pm11.1$\\
 23&$ 24.36\pm 13.89$&$ 0.81\pm 0.08$& 0.95& 12.18& 0.20&11&$166.9\pm10.9$\\
 24&$ 24.84\pm 13.91$&$ 0.82\pm 0.08$& 0.94& 13.04& 0.22&12&$105.2\pm10.0$\\
 25&$ 31.07\pm 11.87$&$ 0.79\pm 0.07$& 0.95& 13.90& 0.24&13&$114.6\pm10.1$\\
\hline
\end{tabular}
\end{table*}

\section{RESULTS} 
\subsection{Hindsight of $A^*_{\rm R} (n)$--$R_{\rm M}(n+1)$ and 
 $A^*_{\rm W} (n)$--$A_{\rm W}(n+1)$ relationships}
In  Table~\ref{table1}  we have given that in intervals $T^*_{\rm M}$ and 
$T^*_{\rm W}$, i.e. 7\,--\,9 months intervals  about one-year after
 the maximum epochs of solar cycles, 
 the sums of the areas of sunspot groups during these intervals,
 $A^*_{\rm R}$ and    $A^*_{\rm W}$
(normalized by 1000)  in  $0^\circ-10^\circ$ latitude intervals of the
southern hemisphere  during Solar Cycles 12\,--\,23 that  have maximal
 correlations with  $R_{\rm M}$ and  $A_{\rm W}$,  respectively, of 
the corresponding next solar cycles~ \citep[also see table~1 in][]{jj21}. 
The values of $A^*_{\rm R}$ and    $A^*_{\rm W}$
 of Solar Cycle 24 that were used for predicting  
$R_{\rm M}$ and  $A_{\rm W}$ of Solar Cycle~25 are also given. 
We made hindsight of the  linear
relationships between $A^*_{\rm R} (n)$   and
 $R_{\rm M} (n+1)$  and  between $A^*_{\rm W} (n)$  and $A_{\rm W} (n+1)$. 
The corresponding details are given in Table~\ref{table2}. 
The hindsight  is reasonably
 good. That is, except in the case of $R_{\rm M}$ of Solar Cycle~18,
in the remaining all cases the correlation is  statistically 
significant at a level above 95\,\% as indicated by Student's t-test.
 In each case the linear-least-square best fit is good, i.e., the slope of 
each linear relation is considerably larger than its uncertainty
 ($\sigma$: standard deviation).
 
Fig.~\ref{f1} shows the comparison of the observed  and the
predicted values of $R_{\rm M}$  and  $A_{\rm W}$  of Solar Cycles
 18\,--\,24. The uncertainties are rms ({\it root-mean-square deviation})
 values.
 In this figure  the predicted values of $R_{\rm M}$ and 
$A_{\rm W}$ of Solar Cycle~25 are also shown. 
As can be seen in this figure in both the cases of  $R_{\rm M}$  and 
 $A_{\rm W}$ there is a reasonably good agreement between the predicted 
and observed values (there exists significant 
correlation between the observed and predicted values). 
 The agreement is much better in the case of $A_{\rm W}$  
than that of $R_{\rm M}$.  The 
 property that the observed  value of $A_{\rm W}$ of Solar Cycle 22 is
 larger than that of
 $A_{\rm W}$ of Cycle~21 is even present in the corresponding 
predicted values of $A_{\rm W}$. Since  here the uncertainties 
(standard errors) in the values of $A_{\rm W}$ are taken care  in the
 calculation of the linear least-square fit between $A^*_{\rm W} (n)$ and
 $A_{\rm W} (n+1)$,  we obtained  slightly higher value, 927 msh,  
for $A_{\rm W}$ of Solar Cycle~25 than that (701 msh) was found in 
\citet{jj21}.

In many solar cycles there is no synchronize 
in the maxima of sunspot number and sunspot area.  
In \citet{jj22} we calculated the linear least-square fit to 
the 13-month smoothed monthly mean values of 
 the area of the sunspot groups in
the Sun's whole sphere (WSGA) and total sunspot number  
(${\rm SN}_{\rm T}$). 
  By using  the predicted value 
of $A_{\rm W}$  from the $A_{\rm W}$--$W_{\rm M}$ relationship 
shown in that paper it was obtained $130\pm12$ for $R_{\rm M}$ of Solar
 Cycle~25 ($W_{\rm M}$ is the maximum value of  13-month smoothed 
monthly mean area of sunspot groups in the Sun's whole sphere 
 during a solar cycle). However, since 
we have used the 13-month smoothed monthly mean values
 throughout the solar cycles,  
 i.e. during maxima, minima, etc. of solar cycles, obviously, there exist 
considerable differences in the  distributions of large and small sunspot
 groups during the solar cycles. It is well-known that the relationship between
  sunspot number and sunspot area is not strictly linear. 
Some scientists have shown that the size distribution of active regions
is close to exponential \citep[e.g.][]{tang84}.
  Some other scientists shown that it is close to power law 
or log-normal distribution \citep{bog88,hz93,howard96}.
 Still some            
scientists have shown that the distribution of sunspot groups 
with respect to maximum area  may not be fitted by a simple one-parameter 
distribution such as single power law or an 
exponential law \citep{gs81}. 
 Fig.~\ref{f2} shows the plot of WSGA versus ${\rm SN}_{\rm T}$.
  As we can see in this figure, obviously the WSGA 
and ${\rm SN}_{\rm T}$  distribution is  not exactly linear. 
The behavior 
of  the  beginning portion that correspond to the  small values of
 WSGA  is somewhat different from that of latter portion that 
correspond to  the large values of WSGA. We calculated  linear 
least-square fit to the logarithm values of WSGA  and  ${\rm SN}_{\rm T}$ 
and  shown in Fig.~\ref{f2}.
 We find  that uncertainty in this fit is considerably lower than that of 
 the corresponding linear fit shown in  \cite{jj22}. A 
value $\approx 1348$ \,msh was obtained for  $A_{\rm W}$ from the 
$A_{\rm W}$--$W_{\rm M}$ relationship (fig.~8 in \cite{jj22}). Here by using
 this value of $A_{\rm W}$ in the relationship
 shown in Fig.~\ref{f2} we obtained  $125\pm11$  for $R_{\rm M}$
 (it is nothing but ${\rm SN}_{\rm T}$ at $T_{\rm M}$) 
 of Solar Cycle~25. It is  slightly smaller than that was predicted earlier.
By using $\approx$927 \,msh of  $A_{\rm W}$ predicted from
 the $A^*_{\rm W} (n)$--$A_{\rm W} (n+1)$ relationship above, we
 obtained $92\pm11$  for $R_{\rm M}$
  of Solar Cycle~25. Both these predicted values are 
also  shown in Fig.~\ref{f2}. The former is slightly 
larger--and the latter is slightly smaller--than the
 observed amplitude of Solar Cycle~24.

\subsection{Prediction for strengths of double peaks of Solar Cycle~25}
\subsubsection{Prediction for the second maximum, $S_{\rm M}$}
Fig.~\ref{f3} shows the variations in the 13-month smoothed monthly
 mean sunspot
 number (SN) during the period 1874\,--\,2017.  In this figure variations
 in the 5-month smoothed monthly SN is also shown.  The 
values of the maximum ($R_{\rm M}$) and the  second largest 
value ($S_{\rm M}$)  
of each of Sunspot Cycles 12\,--\,24 determined from in both these series
 are indicted. In Table~\ref{table3}
 we have given the epochs  $T_{\rm M}$ and $T_{\rm S}$ of 
 $R_{\rm M}$ and  $S_{\rm M}$, respectively, of Sunspot Cycles 12\,--\,24, 
 determined from the 13-month smoothed data. The
intervals (Gnevyshev gaps, in year) between these epochs, 
  the values of the ratios of $S_{\rm M}$ to $R_{\rm M}$, and   the 
 values of the mean and the standard deviation of the corresponding 
absolute values are also given.
As we can see in this table and in Fig.~\ref{f3}, 
 in the case of Solar Cycles 12, 13, 16, 23, and 24 the second 
highest peaks occur first.  The average size of the
Gnevyshev gap is $\approx$1.4-year. In the case of Solar Cycle~19 the gap is
relatively small (only 0.5-year). In fact,  no significant 
 Gnevyshev gap was identified in sunspot data of this cycle 
 \citep[e.g.][]{ng10,rcj21}.
 In the case of Solar Cycles~12 and 24 
the gap is largest, about 2-year. The mean value of the ratio
 $S_{\rm M}/R_{\rm M}$  is 0.9 and the corresponding $\sigma$ is 
 reasonably small.
 That is, the ratio is almost the same in most of the cycles. The ratio is 
somewhat small only in Solar Cycle~15 (there seems to be an ambiguity
 to identify the second highest peak).

Fig.~\ref{f4}(a) shows the correlation between $R_{\rm M}$ and $S_{\rm M}$
 during  Solar Cycles 12\,--\,24 (determined from the values in 
Table~\ref{table3}).  The correlation is reasonably high
 (significant on 99\,\% confidence level). We calculated linear least-square 
fit  by using the Interactive Digital Library (IDL) software
 {\textsf{FITEXY.PRO}}, downloaded from the website
 \textsf{idlastro.gsfcnasa.gov/ftp/pro/math/}.   This software 
 takes into account the  errors in the values of both the  abscissa and  
ordinate  in the calculation of the linear least-square fit.
 Note that a small value of $P$
indicates a poor fit (large $\chi^2$). 
  We obtained the following relationship:
\begin{equation}
S_{\rm M} =  (-9.54 \pm 14.58) + (0.96 \pm 0.08) R_{\rm M}. 
\label{eq1}
\end{equation}
The  least-square best fit is very good, 
i.e. the slope of this linear relationship is about 10 times larger than 
the corresponding $\sigma$. The $\chi^2 = 7.66$ is reasonably small  and the  
corresponding probability (P = 0.74)  is reasonably large. By using this 
 relation and the predicted value $\approx86$ ($\approx 92$)  of
 $R_{\rm M}$ of Solar Cycle~25  we obtain  $73\pm15$
 ($79\pm15$) for
$S_{\rm M}$ of  Solar Cycle~25. The ratio $S_{\rm M}/$$R_{\rm M}$ of 
Solar Cycle~25 is  0.85, which is almost the same as that 
of Solar Cycle~24.

We did hindsight of  the linear
relationships between $R_{\rm M}$   and
 $S_{\rm M}$. The corresponding  details are given in Table~\ref{table4}. 
The hindsight results are reasonably
 good in the sense that except in the case of Solar Cycles~17 and 18,
 in the remaining all cases the correlation is statistically significant and
in each case the best-fit linear relationship is good.
 Fig.~\ref{f4}(b) shows the comparison of the observed  and the
predicted values of $S_{\rm M}$  of Solar Cycles
 17\,--\.24. In this figure  the predicted values of $S_{\rm M}$ of 
 Solar Cycle~25 are also shown.
As can be seen in this figure, except in the case of Solar Cycles~17 and 18, 
 in the remaining all  solar cycles  there is a reasonable good agreement 
between the predicted and the observed values.

In Fig.~\ref{f5} we compare the best-fit  cosine curves of 
  $R_{\rm M}$ \citep[the same 
as shown in fig.~7 of][]{jj22}  and  $S_{\rm M}$ during 
Solar Cycles~12\,--\,24.
The corresponding values of $\chi^2$ are 155 and 104, respectively.
  As we can see in this figure the cosine  best fits of 
 both $R_{\rm M}$  and $S_{\rm M}$  mostly the same (periods are almost equal).
 The extrapolations of these curves yield  $123\pm 33$ 
 for $R_{\rm M}$ and $104\pm28$ for $S_{\rm M}$ of Solar Cycle~25.
 The aforementioned predictions are based on a model where the $\chi^2$ is
large ($>100$) and are thus not particularly reliable.
  A wide range of lengths  (60\,--\,140 years) are suggested for
 Gleissberg cycle \citep[e.g.][]{ogu02}.  The size (143 years)  of the data  
used here is not adequate to determine precisely the long-term 
 periodicity in solar activity. 
 In Fig.~\ref{f5}, there is an indication of 
 the predicted values  of
 $R_{\rm M}$  and $S_{\rm M}$ are at the minimum of upcoming 
 long-period cycle.
 However, this conclusion is not supported by the observations at a
 statistically
significant level (i.e. the null case is not excluded at the 5\,\% level).

\subsubsection{Prediction for first and second peaks (irrespective of heights)}
As we have noticed above in some solar cycles the peak of $R_{\rm M}$ 
occurred first 
and in some other solar cycles the peak of  $S_{\rm M}$  occurred first 
(see Fig.~\ref{f3}, Table~\ref{table3}). In the above analysis (Sec. 3.2.1)  
it is not possible  to predict
  whether the peak of $R_{\rm M}$ 
or that of $S_{\rm M}$ will occur first during the maximum of Solar Cycle~25. 
This is because the peaks of
$R_{\rm M}$ and $S_{\rm M}$ are not in the same
 chronological order in all solar cycles. Therefore,
 the  information on the order of  occurrence 
of  $R_{\rm M}$ and $S_{\rm M}$ in solar cycles  is not given in
 Table~\ref{table3}.
However, it is not required for the purpose of that analysis. 
 We  reorganized the data given in Table~\ref{table3} according to the order 
of occurrence of the peaks that correspond to $R_{\rm M}$ and $S_{\rm M}$.
  Table~\ref{table5} contains the reorganized data, 
i.e. in this table  we gave the epochs  TSN1 and TSN2 of the first
 peak (SNP1) and the second
peak (SNP2), respectively, during the maxima of Sunspot Cycles 12\,--\,24.
It should be noted that both the data of SNP1 and SNP2  contain  
the  values of $R_{\rm M}$  of  some cycles and of $S_{\rm M}$  of some
 other cycles. In Table~\ref{table5} the  values of $R_{\rm M}$  are 
indicated with bold-font.
 The intervals (Gnevyshev gaps, in year) between these
  peaks, i.e TSN2$-$TSN1, the ratios of SNP1 to SNP2, and the values of 
the corresponding mean and standard deviation are also given. As can be 
seen in this table the data of SNP1 contain the  values of $R_{\rm M}$ 
of Solar Cycles 14\,--\,15 and 17\,--\,18 and the values of $S_{\rm M}$
of  Solar Cycles 12, 13 , 16, 23, and 24. Obviously, the data of SNP2
 contain the values of $S_{\rm M}$ of the former cycles and
the values of $R_{\rm M}$ of latter cycles. There is no significant 
difference between the average values SNP1 and SNP2 (almost the same). 
Obviously, the average size of the Gnevyshev gap is the  same as given 
in Table~\ref{table3}. The average value of the ratio SNP1/SNP2 is about one. 
 Solar Cycles 12 and 24 have the same value of the ratio SNP1/SNP2 and
almost the same size  of Gnevyshev gap. In fact, it seems
when SNP2 is larger than SPN1, i.e. when SPN2 represents $R_{\rm M}$,
the corresponding Gnevyshev gap is relatively large, the peaks
are well separated, both peaks are well defined (except in Solar
Cycle~13) and SNP1/SNP2 ratio is to some extent small. In addition,
the corresponding solar cycles might be relatively small (probably
smaller than the respective preceding  solar cycles). All these
 characteristics
also support for a small Solar Cycle~25 and it would have a large
Gnevyshev gap similar to those of Solar Cycles 12 and 24.  
In each hemisphere the temporal behavior of 
the activity in Solar Cycles~24  is almost the same as that of Solar Cycle~12
 and in both of  these solar cycles the peak of whole sphere
activity depict the dominant peak of activity in southern hemisphere 
\citep[see fig.1 in][]{jj20}.  In fact, some authors
 reported that Solar Cycles~12 and 24 are as similar (in shape) 
cycles~\citep{du20}.

 Fig.~\ref{f6} shows the cosine fits to the values of SNP1   and SNP2 during
Solar Cycles~12\,--\,24.
 The corresponding values of $\chi^2$ are 141 and 110, respectively.
  As we can see in this figure the best fit 
cosine  functions  of SNP1 and SNP2 have  periods $\approx$13-cycle and   
$\approx$12-cycle, respectively. That is, the period of SNP1 is about 
one-cycle period (11-year) larger than that of SNP2, and obviously SNP1 leads 
SNP2 by about one year (note that the average size of  Gnevyshev gap 
is about one-year). 
These  results may be somewhat consistent with the  superimposition
  of two waves of solar activity with some phase difference could be a
 cause for the dual-peaks in the maxima  of solar cycles as suggested by  
Gnevyshev~(\citeyear{gne67}, \citeyear{gne77}). However,
 Gnevyshev~(\citeyear{gne67}, \citeyear{gne77}) suggested superimposition 
of  two  $\approx$11-year period waves, whereas the aforementioned result 
 suggests superimposition of two waves of periods $\approx$12-cycle and
 $\approx$13-cycle.   The extrapolations of the
 cosine curves of SNP1 and SNP2  yield  $106\pm34$ 
and  $119\pm28$  for SNP1 and SNP2, respectively, of Solar Cycle~25. 
 These predictions are not particularly reliable because the $\chi^2$
 of the fit is large ($>100$).
 However, form this 
analysis  by a large extent clear that like in Solar Cycle~24, 
in Solar Cycle~25 the second peak would be larger than first peak. 
Obviously, the values of the large and the small  peaks represent 
$R_{\rm M}$ and $S_{\rm M}$, respectively. 
The ratio SNP1/SNP2 of Solar Cycle~25 is about 0.89, which is only slightly
 larger than that of Solar Cycle~24 (see Table~\ref{table5}). 
 In general, all the inferences drawn from the best fit cosine functions  
 have no statistical support, hence  they are at best only 
suggestive rather than compelling.

 Fig.~\ref{f7}(a) shows the correlation between SNP1 and SNP2 during 
Solar Cycle~12\,--\,24. This 
correlation ($r = 0.95$) 
is considerably smaller than that of between $R_{\rm M}$ and $S_{\rm M}$
shown in Fig.~\ref{f4}(a), but still statically significant ($P = 0.05$). 
  We obtained the following relationship between SNP1 and SNP2  by using 
the values of these parameters given in Table~\ref{table5}:
\begin{equation}
{\rm SNP2} =  (31.07 \pm 11.87) + (0.79 \pm 0.07) {\rm SNP1}. 
\label{eq2}
\end{equation}
The  least-square best fit of this relation of SNP1 and SNP2  is 
good, i.e, the slope is about 11 times larger than the 
corresponding standard deviation.
 In this relation  by using the 
 value of SNP1 of Solar Cycle~25 predicted above by extrapolating 
the best-fit cosine curve of SNP1 shown in Fig.~\ref{f5}(a) 
  we get $114.6\pm10.1$ for SNP2 of Solar Cycle~25. It is not
 significantly different from the one predicted from the cosine fit of NSP2.
As we  can see 
in Table~\ref{table6} (after Solar Cycle 18) and in Fig.~\ref{f7}(b)
 the hindsight
 of this relationship suggests a reasonable consistency in the SNP1--SNP2 
relationship and the corresponding prediction is reasonably reliable.  

\subsection{Analysis of 5-month smoothed monthly mean SN}
Since in our earlier analyses we have  predicted 13-month smoothed monthly 
 mean values of the amplitude of Solar Cycle 25, in order to use that predicted
values here (in Sec.~3.2)  we have analysed the 13-month smoothed data of SN.
Some solar cycles contain more peaks during their maxima. We 
 considered only the two peaks which are higher than remaining ones.   
In general there are some solar cycles in which there is a difficulty
 to identify
 Gnevyshev gaps, for example, Solar Cycles 13, 15, and 19  in 13-month 
smoothed monthly mean values of SN. Therefore, here we also analyse
 the data in relatively short intervals: 5-month smoothed monthly
 mean SN data. In this data the Gnevyshev peaks are relatively well 
defined compared to the corresponding peaks  in the 13-month smoothed data
 (see Fig.~\ref{f3}). The epochs of the  peaks during  many solar cycles in the 
 13-month smoothed data closely match with  the corresponding peaks
in the 5-month smoothed data. However,   
there is an ambiguity in determining 
from the 13-month smoothed data the epochs of  $R_{\rm M}$ and 
$S_{\rm M}$ of some solar cycles. For example,
 in the case of Solar Cycles~13 and 15 the positions of the peaks of 
$S_{\rm M}$ in the 13-month smoothed series are seem to be in a large extent
 different  in the 5-month smoothed series. In the case of Solar Cycle~19
 there is peak of $S_{\rm M}$ in the 5-month smoothed data, but it is washed
 out in the 13-month smoothed data (except that there is a slight 
signal of it). In the case of a few solar cycles  $R_{\rm M}$ is first 
and $S_{\rm M}$ is second in the 13-month smooth data, whereas it 
 is opposite in the 5-month smoothed data: for example, Solar Cycles 
13 and 23. In Solar Cycle~23  the values $R_{\rm M}$ and  $S_{\rm M}$ 
are almost equal in the 5-month smoothed data. 

\begin{figure}
\centering
\includegraphics[width=8.5cm]{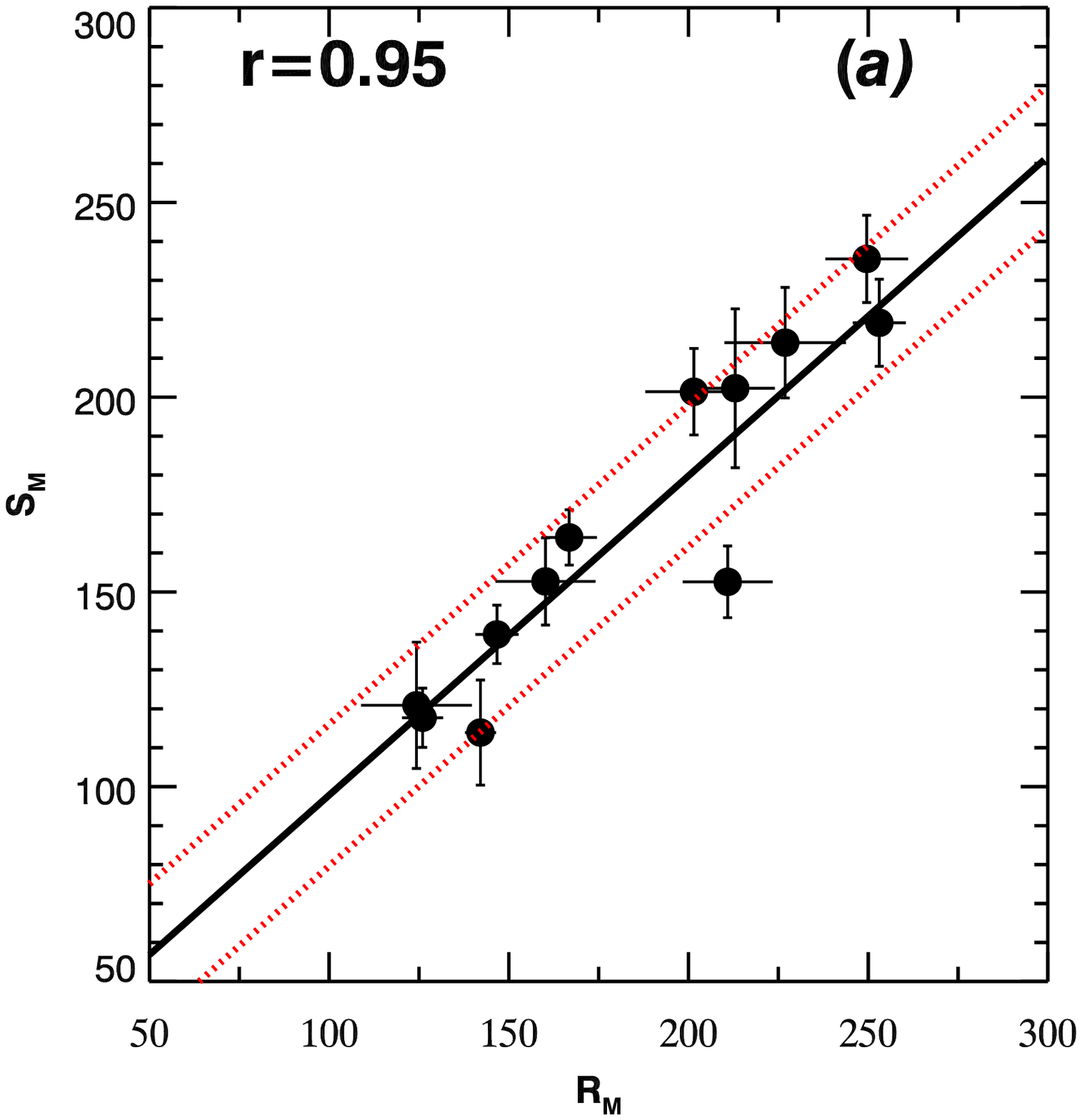}
\includegraphics[width=8.5cm]{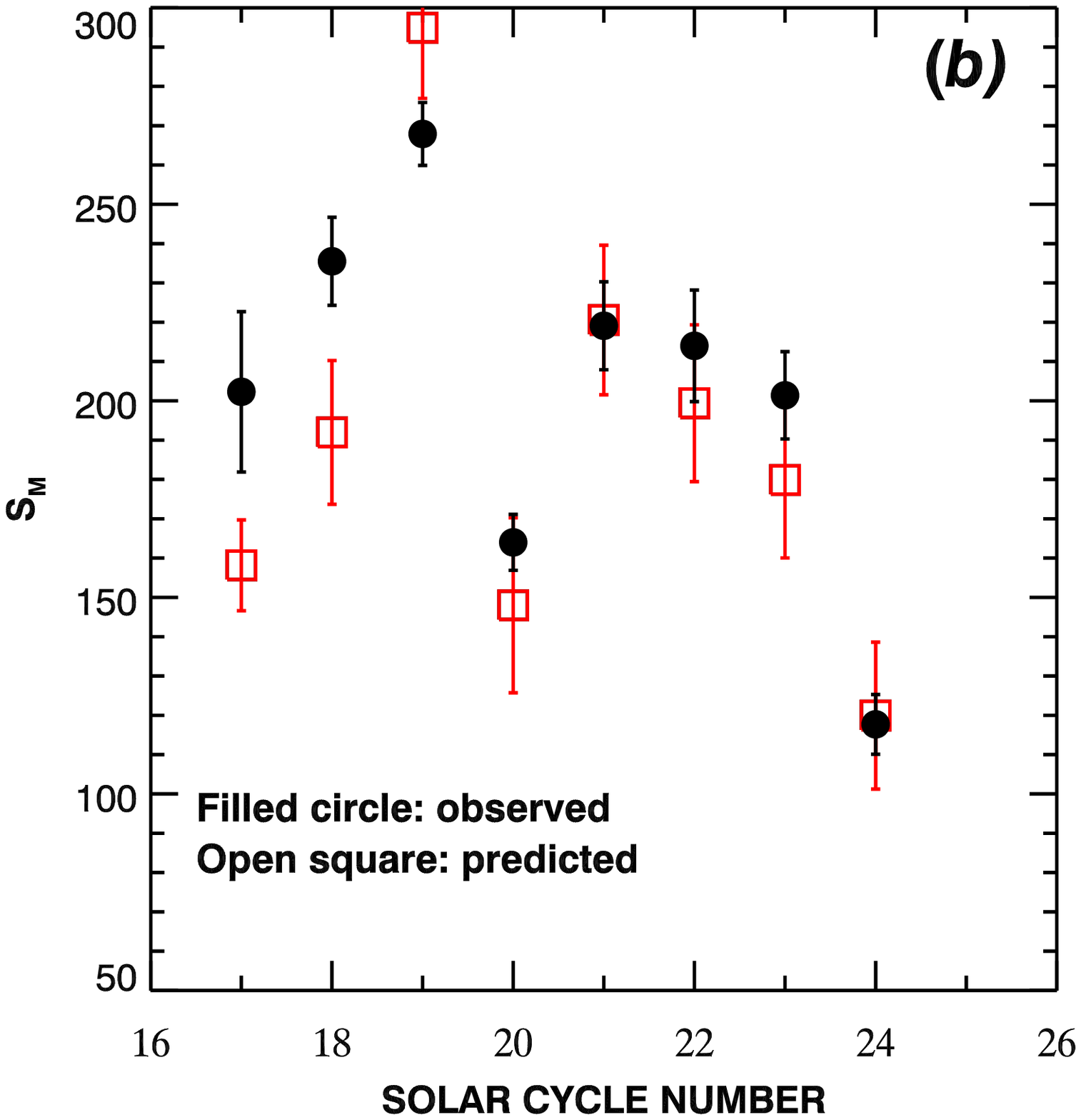}
\caption{({\bf a}) Correlation between $R_{\rm M}$ and $S_{\rm M}$ during
Solar Cycles~12\,--\,24 determined from the 5-month smooth monthly SN.
The {\it continuous line} represents
the best-fit linear relationship, Equation~(3).
The {\it dotted lines} ({\it red}) are drawn at one-rms level.
 ({\bf b}) Hindsight: comparison of the observed and the
predicted values  of $S_{\rm M}$.} 
\label{f8}
\end{figure}

\begin{figure}
\centering
\includegraphics[width=8.5cm]{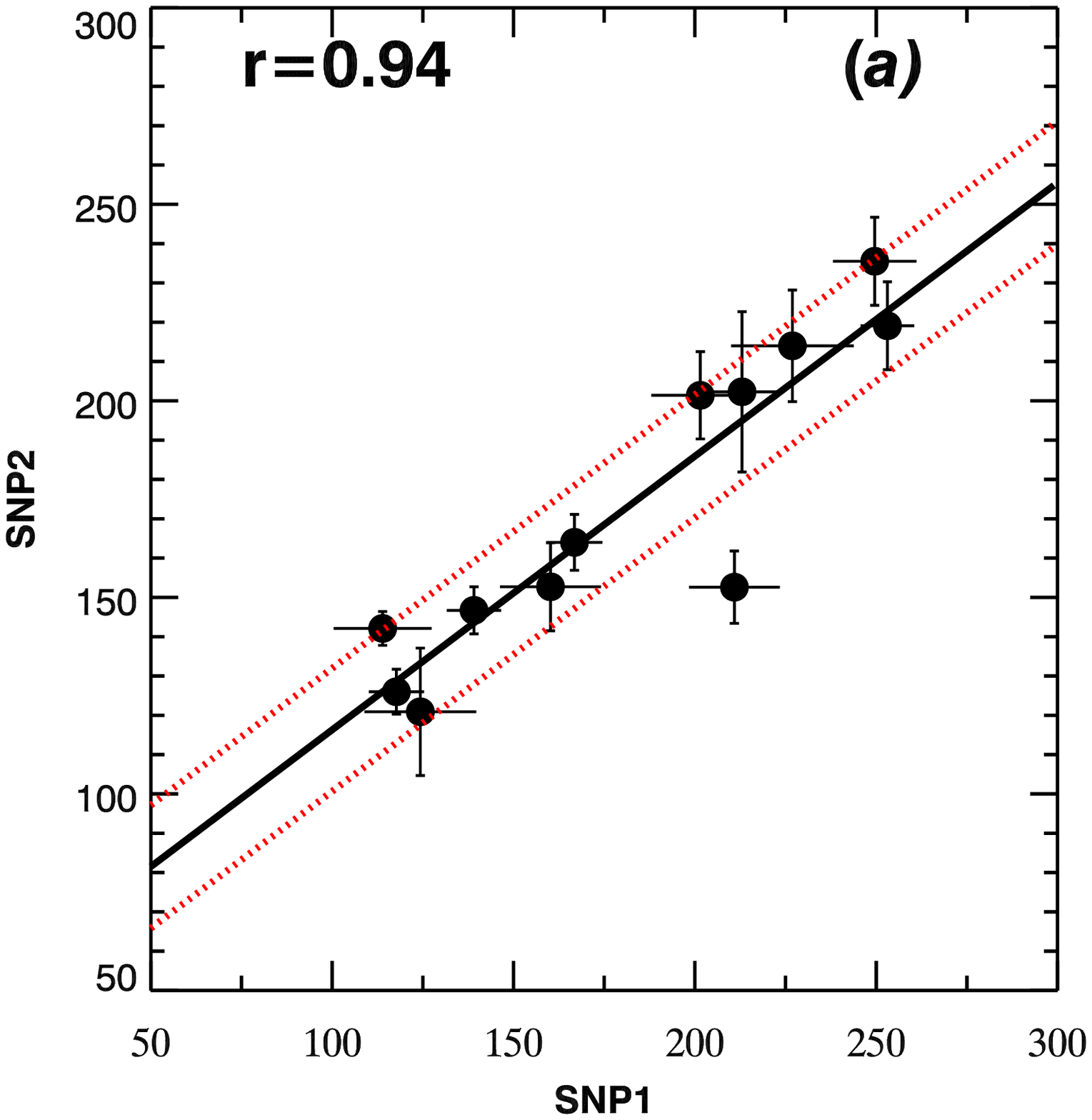}
\includegraphics[width=8.5cm]{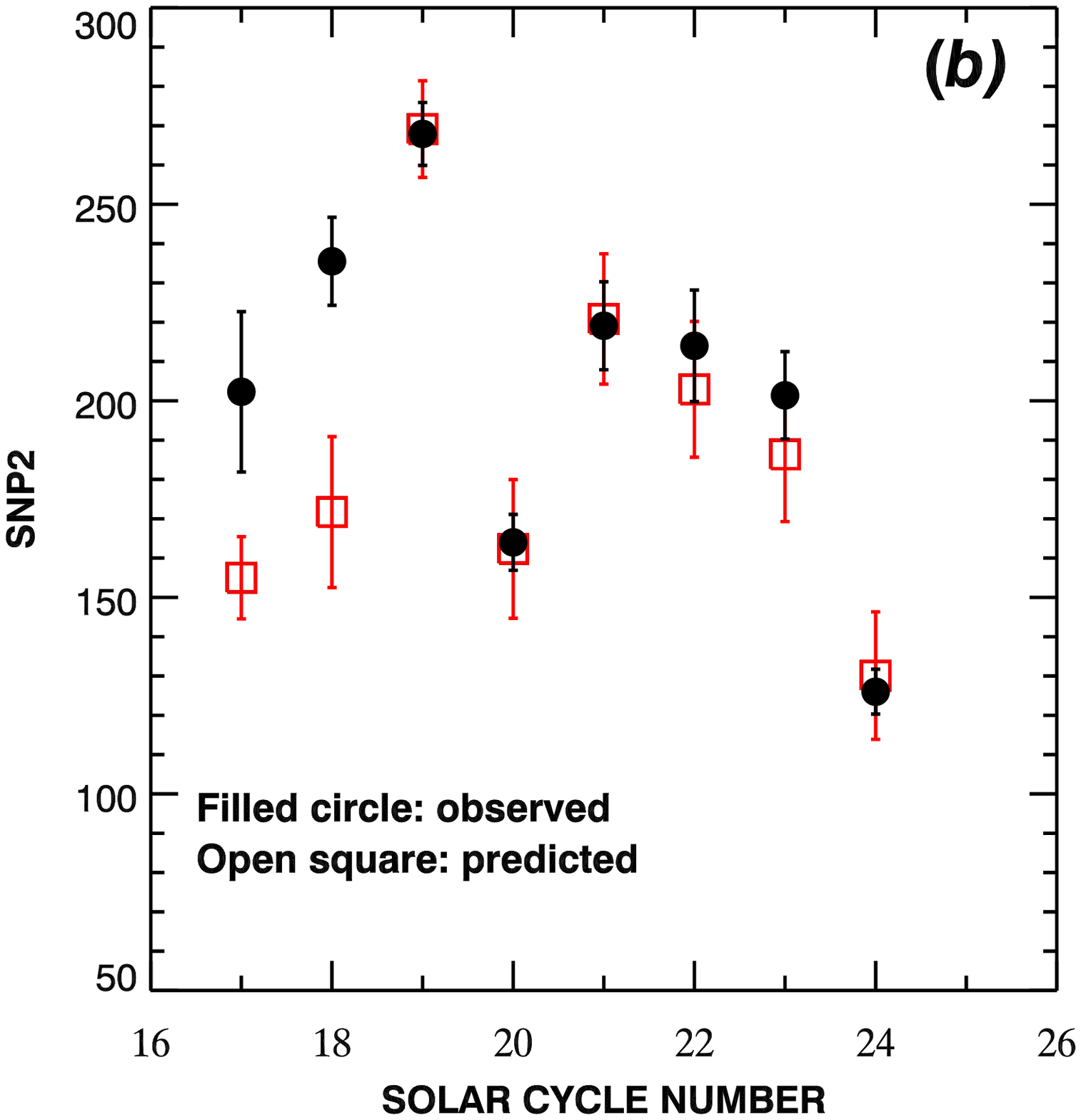}
\caption{({\bf a}) Correlation between SNP1 and SNP2 during
Solar Cycles~12\,--\,24 determined from 5-month smooth monthly SN.
The {\it continuous line} represents
the best-fit linear relationship, Equation~(4).
The {\it dotted lines} ({\it red}) are drawn at one-rms level.
 ({\bf b}) Hindsight: comparison of the observed and the
predicted values  of SNP2.} 
\label{f9}
\end{figure}

 Tables~\ref{table7}, \ref{table8}, \ref{table9}, and \ref{table10}
 are obtained 
from the 5-month smoothed data similarly as Tables~\ref{table3},
\ref{table4}, \ref{table5}, and \ref{table6},
 respectively, that were obtained from the 13-month smoothed data. 
Figs.~\ref{f8} and \ref{f9} are obtained from the 5-month smoothed data
 similarly  as Figs.~\ref{f4} and \ref{f7}, respectively, that were obtained
 from the 13-month smoothed data.  
Obviously, there are considerable 
differences between the sizes of  Gnevyshev gaps of many solar cycles    
determined from the 5-month and 13-month smoothed data, though the 
  the corresponding all cycles'  average sizes  are equal. 
 In Solar Cycles 13 and
  23 the  values of Gnevyshev gaps even have  opposite signs 
(see Tables~\ref{table3} and \ref{table7}). 
There are significant differences in the values of $S_{\rm M}/R_{\rm M}$ of
  solar Cycles 13, 19, and 24  determined from the 5-month and 13-month 
smoothed data. The corresponding 
over all cycles' average values are almost equal.
 Similar arguments  can be made  by  comparing 
 the values of  SNP1 and SNP2  derived from 5-month and 13-month 
smoothed data (see Tables~\ref{table5} and \ref{table9}).

By using the values of $R_{\rm M}$ and $S_{\rm M}$ given in Table~\ref{table7} 
we obtained the following relationship:
\begin{equation}
S_{\rm M} =  (15.86 \pm 13.62) + (0.82 \pm 0.07) R_{\rm M}. 
\label{eq3} 
\end{equation}
The   least-square best-fit of Equation~(\ref{eq3}) by a large extent 
 is good as that 
of Equation~(\ref{eq1}) that derived from the values of 13-month smoothed data.
The parameters of  Equation~(\ref{eq3}) are also given in Table~\ref{table8}.
The slope of this linear relationship is about 11.7 times larger than
the corresponding $\sigma$. The $\chi^2 = 14.8$ is reasonably smaller than 
5\% significant level (i.e. P = 0.19 is much larger than 0.05). 

We obtained the following relationship between SNP1 and SNP2
by using the  values of these parameters given in Table~\ref{table10}:
\begin{equation}
{\rm SNP2} =  (31.07 \pm 11.87) + (0.79 \pm 0.07) {\rm SNP1}. 
\label{eq4}
\end{equation}
The  least-square best fit of this relation of SNP1 and SNP2 is also
 reasonably good. The parameters of Equation~(\ref{eq4}) are also given in
 Table~\ref{table8}. The slope  is about 11.3 times larger than
the corresponding $\sigma$. The $\chi^2 = 16.4$ is to some extent 
smaller than 
5\% significant level (i.e. P = 0.13 is significantly larger than 0.05). 

The hindsight of   
Equations~(\ref{eq3}) and (\ref{eq4}) is shown  in Tables~\ref{table9} 
and \ref{table10} and in Figs.~\ref{f8}(b) and \ref{f9}(b). As can be 
seen in these tables and figures  
 there exists a reasonable consistency in predictions made (for 
Solar Cycle 19\,--\,24)  by using these relations.   
Earlier the 5-month 
smoothed value of $R_{\rm M}$ of Solar Cycle~25 was not predicted.
Hence, here the 5-month smoothed value of $S_{\rm M}$ can not be predicted. 
We did cosine fits to the 5-month smoothed values  of $R_{\rm M}$ and  
$S_{\rm M}$ (not shown here). Although we find the values of $R_{\rm M}$ 
and $S_{\rm M}$ of Solar Cycle~25  are  similar to those obtained
 from the cosine fits shown
 in Fig.~\ref{f5} for 13-month smoothed data,
 the   $\chi^2$ values of the corresponding best fits
 are found to be relatively large.
Hence, here we have not used them. 

Overall, by analyzing  the 5-month smoothed data we confirmed that  
  there is a reasonable consistency in the results derived from the 13-month 
smoothed data. That is, although, obviously,  there are significant differences
 in the Gnevyshev gaps of some solar cycles determined from the 5-month and
 the  13-month smoothed data, they may not have a significant impact on the 
 values of $S_{\rm M}$ predicted above by using the 13-month smoothed data.

\begin{table*}
\caption{The epochs  $T_{\rm M}$ and $T_{\rm S}$  of
  $R_{\rm M}$ and  $S_{\rm M}$, respectively,  of Sunspot Cycles 12\,--\,24 
 determined from the 5-month smoothed monthly mean SN.
 The intervals (Gnevyshev gaps, in year) between these
  peaks, the ratios $S_{\rm M}$/$R_{\rm M}$, and the values of the
 mean and standard deviation of the
 absolute values of these parameters are also given.}
\label{table7}
\begin{tabular}{lcccccccc}
\hline
$n$ &$T_{\rm M}$&$R_{\rm M}$&$\sigma_{\rm M}$&$T_{\rm S}$ &$S_{\rm M}$
&$\sigma_{\rm S}$&$T_{\rm M}-T_{\rm S}$& $S_{\rm M}$/$R_{\rm M}$\\
\hline
 12 &1884.04 &142.1&  4.3& 1882.29& 113.9& 13.5& $-$1.75& 0.80\\
 13 &1893.45 &160.2& 13.9& 1894.45& 152.7& 11.2&    1.00& 0.95\\
 14 &1905.71 &124.3& 15.4& 1907.12& 120.9& 16.2&    1.42& 0.97\\
 15 &1917.54 &210.9& 12.5& 1918.71& 152.6&  9.2&    1.17& 0.72\\
 16 &1928.54 &146.7&  6.0& 1927.12& 139.1&  7.5& $-$1.42& 0.95\\
 17 &1937.45 &213.0& 11.0& 1938.45& 202.3& 20.4&    1.00& 0.95\\
 18 &1947.54 &249.6& 11.5& 1948.46& 235.5& 11.2&    0.92& 0.94\\
 19 &1957.87 &323.5& 13.4& 1958.62& 267.9&  8.0&    0.75& 0.83\\
 20 &1969.29 &166.8&  7.7& 1970.20& 164.0&  7.1&    0.92& 0.98\\
 21 &1979.87 &253.1&  7.4& 1981.71& 219.1& 11.2&    1.83& 0.87\\
 22 &1989.62 &226.9& 16.9& 1991.45& 214.0& 14.2&    1.83& 0.94\\
 23 &2000.37 &201.5& 13.5& 2001.87& 201.4& 11.1&    1.50& 1.00\\
 24 &2014.04 &126.0&  5.7& 2011.87& 117.7&  7.6& $-$2.17& 0.93\\
\hline
Mean&& 195.7& 58.7&& 177.0 & 49.6& 1.34$\pm$0.44&  0.91$\pm$0.08\\
\hline
\end{tabular}
\end{table*}

\begin{table*}
\caption[]{Hindsight: The values of intercept ($C$)
and slope ($D$) of the linear
relationship between $R_{\rm M}$  and
 $S_{\rm M}$ correspond to the predictions for $S_{\rm M}$ of
  Solar Cycles 17\,--\,25 (5-month smoothed monthly values). 
 The corresponding values of the correlation coefficient ($r$),
$\chi^2$ and its probability ($P$), number of data points ($N$),
and predicted values of $S_{\rm M}$ are also given.}
\label{table8}
\begin{tabular}{lccccccccc}
\hline
  \noalign{\smallskip}
$n$ & $C$ & $D$& $r$ & $\chi^2$& $P$&  $N$ & Pred.$S_{\rm M}$ \\
\hline
  \noalign{\smallskip}
 17&$ 81.84\pm 30.27$&$ 0.36\pm 0.19$&  0.74&  3.82& 0.28& 5&$ 158.2\pm 11.6$\\
 18&$ 49.68\pm 38.95$&$ 0.57\pm 0.23$&  0.83&  7.07& 0.13& 6&$ 192.0\pm 18.3$\\
 19&$ -7.18\pm 27.88$&$ 0.93\pm 0.15$&  0.91&  9.18& 0.10& 7&$ 294.9\pm 18.0$\\
 20&$ 12.35\pm 16.73$&$ 0.81\pm 0.08$&  0.95& 10.22& 0.12& 8&$ 148.0\pm 22.3$\\
 21&$ 18.54\pm 16.24$&$ 0.80\pm 0.08$&  0.95& 12.44& 0.09& 9&$ 220.6\pm 19.0$\\
 22&$ 19.00\pm 15.39$&$ 0.80\pm 0.07$&  0.95& 12.45& 0.13&10&$ 199.4\pm 20.0$\\
 23&$ 17.96\pm 15.55$&$ 0.80\pm 0.07$&  0.95& 12.97& 0.16&11&$ 179.9\pm 19.8$\\
 24&$ 17.54\pm 15.77$&$ 0.81\pm 0.08$&  0.94& 14.76& 0.14&12&$ 119.9\pm 18.7$\\
 25&$ 15.86\pm 13.62$&$ 0.82\pm 0.07$&  0.95& 14.80& 0.19&13&   --\\ 
\hline
\end{tabular}
\end{table*}

\begin{table*}
\caption{The epochs  TSN1 and TSN2 of the first peak (SNP1) and the second
peak (SNP2), respectively, of Sunspot Cycles 12\,--\,24 determined from 
the 5-month smoothed monthly mean SN.
 The intervals (Gnevyshev gaps, in year) between these
  peaks, the ratios SNP1/SNP2, and the values of the corresponding
 mean and standard deviation are also given.
The values of $R_{\rm M}$  are indicated with bold-font.}
\label{table9}
\begin{tabular}{lcccccccc}
\hline
$n$ &TSN1&SNP1&$\sigma_1$&TSN2&SNP2&$\sigma_2$&TSN2$-$TSN1& $\frac{\rm SNP1}{\rm SNP2}$\\
\hline
 12& 1882.29& 113.9& 13.5& 1884.04&{\bf 142.1}&{\bf 4.3}& 1.75& 0.80\\
 13& 1893.45&{\bf 160.2}&{\bf 13.9}& 1894.45& 152.7& 11.2& 1.00& 1.05\\
 14& 1905.71&{\bf 124.3}&{\bf 15.4}& 1907.12& 120.9& 16.2& 1.42& 1.03\\
 15& 1917.54&{\bf 210.9}&{\bf 12.5}& 1918.71& 152.6&  9.2& 1.17& 1.38\\
 16& 1927.12& 139.1&  7.5& 1928.54&{\bf 146.7}&{\bf 6.0}& 1.42& 0.95\\
 17& 1937.45&{\bf 213.0}&{\bf 11.0}& 1938.45& 202.3& 20.4& 1.00& 1.05\\
 18& 1947.54&{\bf 249.6}&{\bf 11.5}& 1948.46& 235.5& 11.2& 0.92& 1.06\\
 19& 1957.87&{\bf 323.5}&{\bf 13.4}& 1958.62& 267.9&  8.0& 0.75& 1.21\\
 20& 1969.29&{\bf 166.8}&{\bf  7.7}& 1970.20& 164.0&  7.1& 0.92& 1.02\\
 21& 1979.87&{\bf 253.1}&{\bf  7.4}& 1981.71& 219.1& 11.2& 1.83& 1.16\\
 22& 1989.62&{\bf 226.9}&{\bf 16.9}& 1991.45& 214.0& 14.2& 1.83& 1.06\\
 23& 2000.37&{\bf 201.5}&{\bf 13.5}& 2001.87& 201.4& 11.1& 1.50& 1.00\\
 24& 2011.87& 117.7&  7.6& 2014.04&{\bf 126.0}&{\bf 5.7}& 2.17& 0.93\\
\hline
Mean&& 192.3& 62.64&& 180.4& 45.8& 1.36$\pm$0.44& 1.05$\pm$0.14\\
\hline
\end{tabular}
\end{table*}

\begin{table*}
\caption[]{Hindsight: The values of intercept ($C$)
and slope ($D$) of the linear
relationship between SNP1  and
 SNP2 correspond to the predictions for SNP2 of
  Solar Cycles 17\,--\,25 (5-month smoothed monthly values). 
 The corresponding values of the correlation
coefficient ($r$), $\chi^2$ and its probability ($P$), number of
data points ($N$), and predicted values of SNP2 are also given.}
\label{table10}
\begin{tabular}{lccccccccc}
\hline
  \noalign{\smallskip}
$n$ & $C$ & $D$& $r$ & $\chi^2$& $P$&  $N$ & Pred. SNP2 \\
\hline
  \noalign{\smallskip}
 17&$126.19\pm 16.34$&$ 0.14\pm 0.11$& 0.62&  2.15& 0.54& 5&$ 155.0\pm 10.5$\\
 18&$113.80\pm 19.40$&$ 0.23\pm 0.12$& 0.76&  6.59& 0.16& 6&$ 171.7\pm 19.2$\\
 19&$ 48.78\pm 22.50$&$ 0.68\pm 0.13$& 0.87& 14.86& 0.01& 7&$ 269.1\pm 12.3$\\
 20&$ 49.67\pm 13.75$&$ 0.68\pm 0.07$& 0.94& 14.86& 0.02& 8&$ 162.3\pm 17.6$\\
 21&$ 50.22\pm 13.35$&$ 0.67\pm 0.07$& 0.94& 14.89& 0.04& 9&$ 220.8\pm 16.6$\\
 22&$ 50.59\pm 12.82$&$ 0.67\pm 0.06$& 0.94& 14.90& 0.06&10&$ 202.9\pm 17.3$\\
 23&$ 49.94\pm 12.92$&$ 0.68\pm 0.06$& 0.94& 15.25& 0.08&11&$ 186.4\pm 17.1$\\
 24&$ 49.70\pm 12.95$&$ 0.68\pm 0.06$& 0.94& 16.27& 0.09&12&$ 130.1\pm 16.2$\\
 25&$ 46.72\pm 11.26$&$ 0.70\pm 0.06$& 0.94& 16.45& 0.13&13&  --\\      
\hline
\end{tabular}
\end{table*}

\begin{table}
\centering
\caption{DM (in $\mu$Tesla) and $\sigma_{\rm DM}$ (in $\mu$Tes) represent 
 the average 
dipole moment in the 3-year interval $T^*_{\rm DM}$ just
 before the end of a solar cycle and the corresponding uncertainty, 
respectively, determined from Wilcox Observatory polar fields data for Solar
 Cycles 21\,--\,24 and  it is taken from the paper  by
Jiang et. al (2007)  for 
Solar Cycle 20  ($\sigma_{\rm DM}$ is not available), 
which     was determined   from MWO polar fields data by 
Svalgaard, Cliver, \& Kamide (2005). The symbol $^\mathrm{a}$ 
indicates the average  $\sigma_{\rm DM}$ of Cycles 21\,--\,24.}
\label{table11}
\begin{tabular}{lccc}
\hline
  \noalign{\smallskip}
$n$ &$T^*_{\rm DM}$&DM&$ \sigma_{\rm DM}$ \\
  \noalign{\smallskip}
\hline
  \noalign{\smallskip}
20&1973.21-1976.21&250&1.4$^\mathrm{a}$\\
21&1983.71-1986.71&247.8&2.7\\
22&1993.62-1996.62&200.3&1.2\\
23&2005.96-2008.96&112.9&0.9\\
24&2016.96-2019.96&125.8&0.8\\
\hline
  \noalign{\smallskip}
\end{tabular}
\end{table}

\begin{figure}
\centering
\includegraphics[width=10cm]{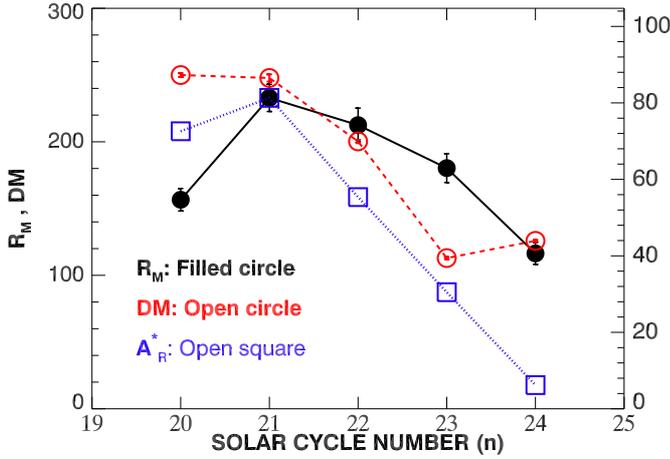}
\caption{Variations in average  dipole moment (DM)  over last 3 years
of solar cycles,  the sum  ($A^*_{\rm R}$) of the areas of sunspot groups in
$0^\circ - 10^\circ$ latitude during a small (7 months) interval
just after the maxima of solar cycles, and  the amplitude
($R_{\rm M}$, maximum yearly mean value of sunspot number) of
 the solar cycle.  In the case of $R_{\rm M}$  and DM the error bars are
 $1\sigma$ (standard deviation) levels (the error in a value of DM is
very small).} 
\label{f10}
\end{figure}

\begin{figure}
\centering
\includegraphics[width=8.5cm]{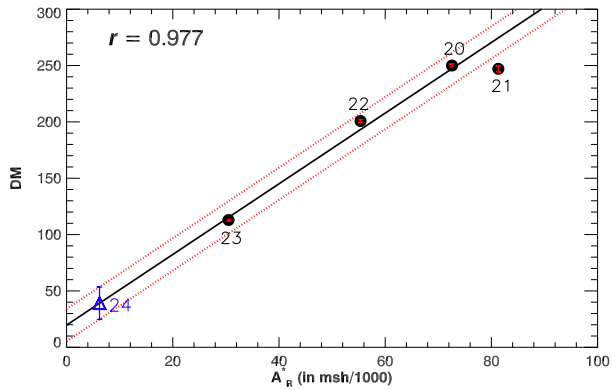}
\caption{The scatter plot of $A^*_{\rm R}$ versus DM of 
Solar Cycles 20\,--\,23. 
The {\it continuous line} represents the best-fit linear relation, 
 Equation~(5)  and {\it dotted curves} represent one-rms. The corresponding 
value of the correlation-coefficient ($r$) is given and Waldmeier solar 
solar cycle number is also shown. The {\it triangle} ({\it blue}) represents
 the derived 
strength of the average DM of 3 years  before the end of Solar Cycle 24.} 
\label{f11}
\end{figure}

\begin{figure}
\centering
\includegraphics[width=8.5cm]{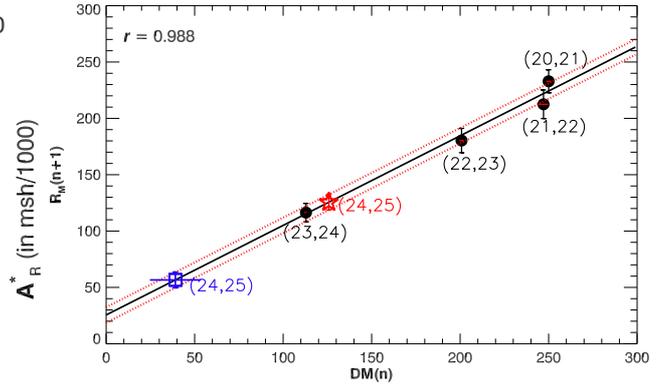}
\caption{Plot of DM ($n$) versus $R_{\rm M}$ ($n+1$), where $n = 20,\dots,23$
 represents the Waldmeier solar cycle number. 
The {\it continuous line} represents the best fit  linear relationship,
 Equation~(6)
 and the {\it dotted curves} represent one-rms level. The corresponding 
value of $r$ is given and the pairs of Waldmeier
 solar solar cycle numbers are also shown. The symbol {\it star} ({\it red})
  represents 
the $R_{\rm M}$ of Solar Cycle~25 predicted by substituting in Equation~(6) 
the observed mean value  of DM. The symbol {\it square} ({\it blue})
 represents the value   that obtained by substituting  
 in  Equation~(6) the predicted  average value of DM over last 3-years of
 Solar Cycle 24.} 
\label{f12}
\end{figure}

\begin{figure}
\centering
\includegraphics[width=8.5cm]{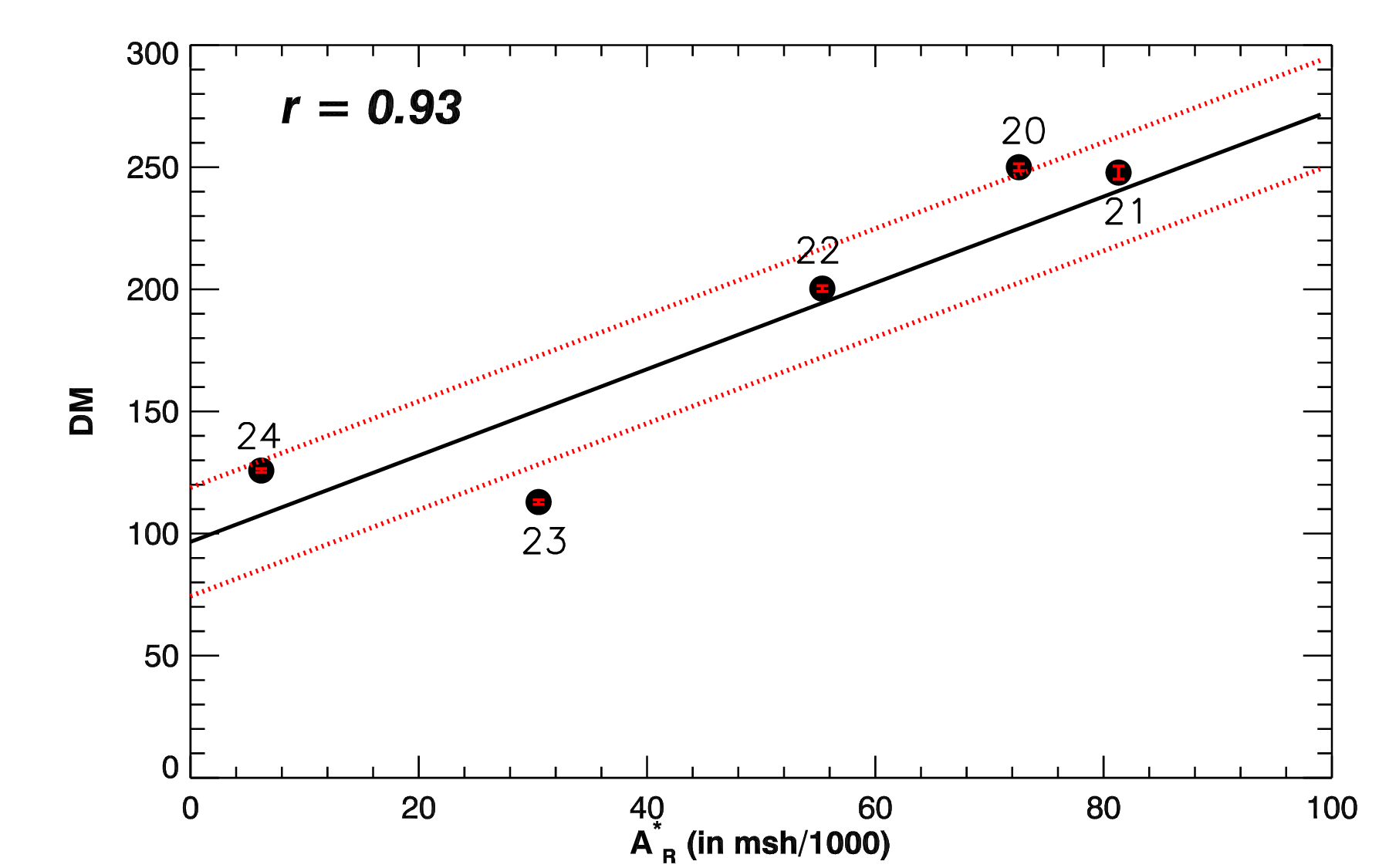}
\caption{The scatter plot of
 $A^*_{\rm R}$ versus DM of Solar Cycles 20\,--\,24. 
The {\it continuous line} represents the best-fit linear relation 
 and the {\it dotted curves}  represent one-rms level. The corresponding 
value of $r$ is given and Waldmeier solar 
solar cycle number is also shown.
This figure is the same as Fig.~\ref{f9}, but the observed value of DM
 of Solar Cycle 24 is included.}
\label{f13}
\end{figure}

\subsection{Comparison between $A^*_{\rm R}$  and DM}
In Table~\ref{table11} we have given the values of DM of Solar Cycles 20--24.
Fig.~\ref{f10} shows the cycle-to-cycle variations in 
 $R_{\rm M}$,  $A^*_{\rm R}$, and DM during Solar Cycles 20--24 (the error 
in  DM is very small).  As can be seen in this figure 
 the profiles of all these parameters are closely similar. 
However, the pattern of DM of all the five solar cycles, 20--24, 
 is  somewhat different.  $A^*_{\rm R}$ is considerably 
decreased from Solar Cycle~23 to Solar Cycle 24. In fact, 
  $A^*_{\rm R}$  
 monotonically decreased from Solar Cycle 21 to Solar Cycle  24.
DM  also decreased from Solar Cycle 21 to Solar Cycle~23, but  slightly 
increased from Solar Cycle 23 to Solar Cycle 24. 

Fig.~\ref{f11} shows the scatter plot of 
 $A^*_{\rm R}$  versus DM during Solar Cycle 20\,--\,23. The corresponding 
correlation ($r = 0.98$) is very good, $i.e.$ it is
 statistically highly significant
 (Student's $t=6.5$,  $t = 4.3$ for $p = 0.05$  for 2 degree of 
freedom). We obtained the 
following linear relationship between
  $A^*_{\rm R}$ and DM during Solar Cycles 20\,--\,23:
\begin{equation}
{\rm DM} =  (19.7 \pm 1.7) + (3.1 \pm 0.3 ) A^*_{\rm R}. 
\label{eq5}
\end{equation}
The uncertainties (see Table~\ref{table11}) in the value  of DM are taken care 
in the least-square fit calculations. The  best-fit linear relation
 is reasonably good, 
$i.e.$ the  slope is about ten times larger than the corresponding $\sigma$.
 $\chi^2 =3.1$ is reasonably small
 (note that $\chi^2 =7.815$ for $p = 0.05$ for 3 degree of freedom). 
Except the data point of  Solar Cycle~21, the remaining  
 three data points are laying within one-rms level.
 In Equation~(\ref{eq5}) by substituting 
 the value of  $A^*_{\rm R}$ (given in Table~\ref{table1}) of Solar Cycle~24,
 we obtained $39\pm14$ for DM of Solar Cycle 24 (${\rm rms} =14$
 looks to be  relatively large, but 
this predicted value of DM is close to the lower end of the large 
range of DM values). This predicted value 
of DM of Solar Cycle~24  is also shown in Fig.~\ref{f11}.

Fig.~\ref{f12} shows the scatter plot of DM(n) versus $R_{\rm M}$ (n+1), 
where $n = 20,\dots,23$ represents the Waldmeier solar cycle number. 
The corresponding correlation  ($r = 0. 99$) is statistically highly
 significant 
(Student's $t = 9.1$). We obtained the following linear relationship:
\begin{equation}
 R_{\rm M} (n+1) =  (25 \pm 18) + (0.79 \pm 0.08){\rm DM}(n).
\label{eq6}
\end{equation}
 The uncertainties of both DM and $R_{\rm M}$ are taken care in 
this linear least-square fit calculations. 
The  least-square fit to the data is reasonably good, $i.e.$ 
  $\chi^2 = 1.47$ is  small and the corresponding  $P = 0.48$ (the
  $\chi^2$ is considerably smaller than 5\,\% confidence level).   
 The  rms$=6.9$  is also considerably small 
and almost all the  data points are within the one-rms level.
In  Equation~(\ref{eq6}) by substituting the predicted 
and observed  (see Table~\ref{table11}) values of DM of Solar Cycle~24,
 we obtained 
the values $57\pm 7$ and $125 \pm 7$, respectively, for $R_{\rm M}$ of  
 Solar Cycle 25. These predicted values are also sown in Fig.~\ref{f12}.
The latter and the value predicted from WSGA--$\rm {SN}_{\rm T}$ relationship
(shown in Fig.~\ref{f2}) above, are agree each other very well. However,
the value ($86 \pm 18$) is predicted for $R_{\rm M}$ of Solar Cycle~25
in \cite{jj21}   by using $A^*_{\rm R} (n)$--$R_{\rm M} (n+1)$ linear 
relationship  is much higher than the former and considerably lower
 than the latter.
The predicted value for $R_{\rm M}$ of Solar Cycle 25 by using 
the observed value of DM is  substantially  
(about 119\,\%)
larger than that predicted by using the predicted value of DM. 
The former is slightly larger--whereas the latter is 
substantially lower--than the value of $R_{\rm M}$  of Solar Cycle~24.

Since the corresponding  correlations of both the
 $A^*_{\rm R} (n)$ -- $R_{\rm M} (n+1)$ and 
  ${\rm DM} (n)$--$R_{\rm M} (n+1)$ (Equation~(\ref{f6})) 
 linear relationships are  high, hence we can
 expect a reasonable  high correlation between
$A^*_{\rm R}$ and DM. Fig.~\ref{f13} shows the correlation
 between $A^*_{\rm R}$ and 
DM determined from the values of all the five pairs of data of Solar 
 Cycles 20\,--\,24. 
 The correlation ($r = 0. 93$) is  larger 
 than that of 5\,\% significant 
level (Student's $t = 4.5$), but substantially lower than that determined 
from the four pairs of data of Solar Cycles 20\,--\,23 shown in Fig.~\ref{f11}. 
$\chi^2 = 15.7$  is much larger then that $\chi^2 = 9.488$ of 5\,\% 
significant level for four degrees of freedom
  and ${\rm rms} = 22.2$ is also  relatively large. That is, in this case  
there is a relatively large scatter in the data points.    

Overall, the value of DM predicted for Solar Cycle~24 is much smaller
 than the observed one (see Table~\ref{table11}). Obviously, the predicted 
value of DM is
 incorrect. Therefore, the predicted  value of $R_{\rm M}$ of 
  Solar Cycle~25 by using the predicted value of DM  of 
Solar Cycle~24 is also incorrect. In addition, the correlation between 
$A^*_{\rm R}$ and DM determined from the values of all the five pairs of
 data of Solar Cycles 20\,--\,24 is weak. All these imply that there exists 
only a weak  relationship between $A^*_{\rm R}$ and DM in Solar Cycle 24.

\section{DISCUSSION AND CONCLUSIONS}
In a series of papers, we predicted the amplitudes of  Solar Cycles 24 
and 25 by using the  linear relationship between  $A^*_{\rm R}$  of
a solar cycle (n)  and $R_{\rm M}$  of the next solar cycle (n+1).
In the present analysis  by verifying the
 $A^*_{\rm R} (n)$--$R_{\rm M} (n+1)$ and 
$A^*_{\rm W} (n)$--$A_{\rm W} (n+1)$ 
relationships through hindsight we  confirmed that there is a good 
 consistency in this method of  prediction for the  amplitude of
 a solar cycle.  From this method a value $86\pm 18$ ($92\pm11$) is 
 predicted for $R_{\rm M}$ of Solar Cycle~25 \citep{jj21}. 
Recently, by fitting a cosine function  to the 
cycle-to-cycle modulations in the maxima of the  mean area of
 sunspot groups of Solar Cycles 12\,--\,24 and using the existence 
of a reasonably good linear
 relationship between the long-term variations of sunspot-group area
 and sunspot number we  predicted $130\pm 12$ for  $R_{\rm M}$ of
 Solar Cycle~25 \citep{jj22}.
In the present analysis
we have made an improvement in the relationship between long-term variations
of  sunspot number and sunspot-group area, Therefore, the aforementioned
 prediction is found  to be $125\pm11$.
We  show the existence of a good correlation between the 
 strength of
 polar fields (DM) at the end of  a solar cycle $n$ and  the amplitude
($R_{\rm M}$)  of
solar  cycle $n+1$.  We predicted  $R_{\rm M}$ 
 of  Solar Cycle~25
by  using the strength of  polar fields (DM) at the end of Solar Cycle 24. 
We found $125\pm7$ for $R_{\rm M}$ of Solar Cycle~25. 
 This and  the  value $125\pm11$ predicted from the aforementioned 
 previous method agree each other very well, but  considerably 
larger than the  value predicted  by using  the 
 $A^*_{\rm R} (n)$--$R_{\rm M} (n+1)$ relationship.

 We find that there exits a good correlation between 
$R_{\rm M}$ and $S_{\rm M}$ during the  Solar Cycles~12\,--\,24. 
 By using the predicted value $\approx 86$ ($\approx 92$) of $R_{\rm M}$
 of Solar Cycle~25 and the $R_{\rm M}$--$S_{\rm M}$ linear relation 
we predict $73\pm15$ ($79\pm15$) for $S_{\rm M}$ of Solar Cycle~25. 
The value 0.85 of the
 ratio $S_{\rm M}$/$R_{\rm M}$ of Solar Cycle 25 is found to be almost 
the same  as that of Solar Cycle~24.
The  cosine fits to the values of the first and the second peaks 
(irrespective of their heights) 
 of Solar Cycles~12\,--\,24 suggest  the existence of  
 $\approx$13-cycle and $\approx$12-cycle periods in the variations of 
the first and second peak
values, respectively. Moreover, from this analysis 
we find that in Solar Cycle~25 $S_{\rm M}$  would 
occur before $R_{\rm M}$, the  same as in Solar Cycle~24.
However, this analysis  suggests $\approx$106 and  $\approx$119
 for $S_{\rm M}$ and $R_{\rm M}$  of Solar Cycle 25, respectively.
 Since in our earlier analyses we have  predicted 13-month smoothed monthly
 mean values of the amplitude of Solar Cycle 25, in order to use them here 
 we have analysed the 13-month smoothed data of SN to determine the Gnevyshev 
gaps. However, through the analysis of the data in relatively small 
interval (the 5-month smoothed monthly SN),  we 
confirmed that  there is a reasonable consistency in the results derived from
 the 13-month smoothed data. 

 A good correlation between ${\rm DM} (n)$ and $R_{\rm M} (n+1)$, 
 that too from a few
 pairs of data points, may be not sufficient to make a reliable
 prediction. However, 
 this method has a  support from  a kind of magnetic flux-transport dynamo
 models \citep{jiang07,kumar21}.
Since the corresponding  correlations of both the   
 $A^*_{\rm R} (n)$--$R_{\rm M} (n+1)$ and   
 ${\rm DM} (n)$--$R_{\rm M} (n+1)$ relationships are high, 
hence one can expect a high 
correlation between  $A^*_{\rm R}$ and DM of a solar cycle, so that in 
principle  by  using 
$A^*_{\rm R}$ of  a solar cycle  DM of the solar cycle  can 
 be predicted by about 3 years in advance.
However, the value ($39\pm14$) of DM of Solar Cycle~24 that predicted 
by using 
the reasonably good correlation between  $A^*_{\rm R}$ and DM during 
Solar Cycles 20\,--\,23 is found to be much smaller than the corresponding 
observed value (see Table~\ref{table7}).
Obviously, the predicted value of DM is incorrect. 
$A^*_{\rm R}$  monotonically decreased from Solar Cycle 21 to
 Solar Cycle  24. DM also decreased from Solar Cycle 21 to Solar Cycle~23,
but  slightly increased from Solar Cycle 23 to Solar Cycle 24, so that
  the correlation between DM and $A^*_{\rm R}$ during
 Solar Cycles~20\,--\,24 is found to be to some extent weak. All 
these suggest that the relationship (if exists) between $A^*_{\rm R}$ and DM  
is weak. 

 The epoch of $A^*_{\rm R}$ of a solar cycle is close to the epoch of 
  change in the polarity of  global magnetic field.  
Hence, $A^*_{\rm R}$ is related to
 emergence of new magnetic flux/cancellation of old flux, globally.
Therefore,  the existence of a good 
correlation between  $A^*_{\rm R}$ and DM  may be connected   to the
global evolution of the solar magnetic fields during the declining phase 
of the solar cycle. 

In the present analysis
 we cannot conclude which one of the predictions for the amplitude of 
Solar Cycle~25 mentioned above,  
will be correct. The predictions made by the cosine fits of sunspot
 data agrees well with  the
prediction based on the strength of polar fields. However, the cosine fits 
have large uncertainties (the values of $\chi^2$  are to some extent large). 
 Here we  find that there is a good 
consistency  in the $A^*_{\rm R} (n)$--$R_{\rm M}(n+1)$ relationship. Hence, 
 we may able to claim that our prediction based on this relationship 
 is reasonably reliable.   

\section*{acknowledgements}
 The author thanks the anonymous reviewer for  useful comments
 and suggestions. The author also thanks Luca Bertello for valuable
suggestions. 
The author acknowledges the work of all the
 people contribute  and maintain the GPR and DPD  sunspot databases and 
 the polar-fields data  measured in WCO.
The sunspot-number data are provided by WDC-SILSO, Royal Observatory of
Belgium, Brussels. 

\section*{data Availability}
 All data generated or analysed during this study are included in this
 published article.

{}
\bsp   
\label {lastpage}

\begin{thebibliography}{}
\bibitem[\protect\citeauthoryear{Bazilevskaya \etal}{2000}]{baz00}
Bazilevskaya, G.A., Krainev, M.B., Makhmutov, V.S., F;l\"ukiger, E.O. 
Sladkova, A.I., Storini, M. 2000, \solphys, 197, 157 
\bibitem[\protect\citeauthoryear{Bhowmik \& Nandy}{2018}]{bn18}
Bhowmik, P., Nandy, D., 2018,  Nat. Comm., 9, A5209
\bibitem[\protect\citeauthoryear{Bogdan \etal}{1988}]{bog88}
Bogdan, T.J., Gilman, P.A., Lerche, I., Howard, R., 1988, \apj, 327, 451
\bibitem[\protect\citeauthoryear{Cameron, Jiang, \& Sch\"ussler}{2016}]{cameron16}
Cameron, R.H, Jiang, J., Sch\"ussler, M., 2016, \apjl, 823, 122
\bibitem[\protect\citeauthoryear{{Clette  \& Lef\'vre}}{2016}]{clette16}
Clette, F., Lef\'evre, L., 2016, \solphys, 291, 2629 
\bibitem[\protect\citeauthoryear{{Dikpati \& Gilman}}{2006}]{dg06}
Dikpati, M., Gilman, P.A., 2006, \apj, 649, 498
\bibitem[\protect\citeauthoryear{{Du}}{2015}]{du15}
Du, Z.L., 2015, \apj, 804, 3 
\bibitem[\protect\citeauthoryear{{Du}}{2020}]{du20}
Du, Z.L., 2020, \solphys, 295, 134.
\bibitem[\protect\citeauthoryear{Feminella \& Storini}{1997}]{fs97}
Feminella, F., Storini, M. 1997, \aap, 322, 311
\bibitem[\protect\citeauthoryear{Gnevyshev}{1967}]{gne67}
Gnevyshev, M.N., 1967,  \solphys, 1, 107
\bibitem[\protect\citeauthoryear{Gnevyshev}{1977}]{gne77}
Gnevyshev, M.N., 1977, \solphys, 51, 175
\bibitem[\protect\citeauthoryear{{Gokhale \& Sibaraman}}{1981}]{gs81}
Gokhale, M.H., Sivaraman, K.R., 1981, \japa, 2, 365
\bibitem[\protect\citeauthoryear{{Gonzalez, Gonzalez, \& Tsurutani}}{1990}]{ggt90}
Gonzalez, W.D., Gonzalez, A.L.C., Tsurutani, B.T., 1990, \planss, 38, 181
\bibitem[\protect\citeauthoryear{{Hathaway \& Upton}}{2016}]{hath16}
Hathaway, D.H., Upton, L.A., 2016, \jgr, 121, 10744
\bibitem[\protect\citeauthoryear{Harvey \& Zwaan}{1993}]{hz93}
Harvey, K.L., Zwaan, C., 1993, \solphys, 148, 85
\bibitem[\protect\citeauthoryear{Howard}{1996}]{howard96}
Howard, R., 1996, \araa, 34, 75
\bibitem[\protect\citeauthoryear{Javaraiah}{2007}]{jj07}
Javaraiah, J., 2007, \mnras,  377, L34 
\bibitem[\protect\citeauthoryear{Javaraiah}{2008}]{jj08}
Javaraiah, J., 2008, \solphys, 252, 419 
\bibitem[\protect\citeauthoryear{Javaraiah}{2015}]{jj15}
Javaraiah, J., 2015, \na, 34, 54
\bibitem[\protect\citeauthoryear{Javaraiah}{2020}]{jj20}
Javaraiah, J., 2020, \solphys, 295, 8 
\bibitem[\protect\citeauthoryear{Javaraiah}{2021}]{jj21}
Javaraiah, J., 2021, \apss, 366, 16 
\bibitem[\protect\citeauthoryear{Javaraiah}{2022}]{jj22}
Javaraiah, J., 2022, \solphys, 297, 33 
\bibitem[\protect\citeauthoryear{Jiang, Chatterjee, \& Choudhuri}{2007}]{jiang07}
Jiang, J., Chatterjee, P., Choudhuri, A.R., 2007, \mnras, 381, 1527
\bibitem[\protect\citeauthoryear{Kilcik \& Ozg\"uc}{2014}]{koz14}
Kilcik, A., Ozg\"uc, A., 2014, \solphys, 289, 1379
\bibitem[\protect\citeauthoryear{Kumar \etal}{2021}]{kumar21}
Kumar, P., Nagy, M., Lemerle, A., Karak, B.B., Petrovay, K., 2021,
 \apj, 909, 87
\bibitem[\protect\citeauthoryear{Norton \& Gallagher}{2010}]{ng10}
Norton, A.A.,  Gallagher, J. C., 2010, \solphys, 261, 193
\bibitem[\protect\citeauthoryear{Ogurtsov \etal}{2002}]{ogu02}
 Ogurtsov, M.G., Nagovitsyn, YU.A, Kocharov, G.E., Jungner, H., 2002, 
\solphys, 211, 371 
\bibitem[\protect\citeauthoryear{Pandey, Hiremath, \& Yellaiah}{2017}]{phy17}
Pandey, K.K,  Hiremath, K.M., Yellaiah, G., 2017, \apss, 362, 106 
\bibitem[\protect\citeauthoryear{Pesnell}{2008}]{pes08}
Pesnell, W.D., 2008, \solphys, 252, 209
\bibitem[\protect\citeauthoryear{Pesnell}{2018}]{pesnell18}
Pesnell, W.D., 2018, Space Weather, 16, 1997 
\bibitem[\protect\citeauthoryear{Ravindra \& Javaraiah}{2015}]{rj15}
Ravindra, B.,  Javaraiah, J., 2015, \na, 39, 55
\bibitem[\protect\citeauthoryear{Ravindra, Chowdhury, \&  Javaraiah}{2021}]
{rcj21}
Ravindra, B., Chowdhury, P.,  Javaraiah, J., 2021, \solphys, 296, 2 
\bibitem[\protect\citeauthoryear{Schtten \etal}{1978}]{sch78}
Schatten, K.H., Scherrer, P.H., Svalgaard, L.,  Wilcox, J.M. 1978,
\grl, 5, 411.
\bibitem[\protect\citeauthoryear{Storini \etal}{1997}]{stori97}
Storini, M., Pase, S., S\'ykora, J., Parisi, M., 1997, \solphys, 172, 317
\bibitem[\protect\citeauthoryear{Storini \etal}{2003}]{sto03}
Storini, M., Bazilevskaya, G.A., Fl\"ukiger, E.O., Krainev, M.B.,
 Makhmutov, V.S., Sladkova, A.I., 2003, Adv. Space Res., 31, 895
\bibitem[\protect\citeauthoryear{Svalgaard, Cliver, \& Kamide}{2005}]{sval05}
Svalgaard, L., Cliver, E.W., Kamide, Y., 2005, \grl, 32, L01104
\bibitem[\protect\citeauthoryear{Tang, Howard, \& Adkins}{1984}]{tang84}
Tang, F., Howard, R., Adkins, J.M., 1984, \solphys, 184, 41 
\bibitem[\protect\citeauthoryear{Temmer \etal}{2006}]{temm06}
Temmer, M., Ryb\'ak, J., Bend\'ik, P., Veronig, A., Vogler, F., Otruba,W.,
 P\"otzi,W., Hanslmeier, A.: 2006, \aap, 447, 735
\bibitem[\protect\citeauthoryear{Upton \& Hathaway}{2018}]{uh18}
Upton, L.A., Hathaway, D.H., 2018, \grl, 45, 8091
\bibitem[\protect\citeauthoryear{Wang}{2017}]{wang17}
Wang, Y.-M., 2017, \ssr, 210, 351
\end{thebibliography}
\end{document}